\documentclass[a4paper,11pt]{article}
\pdfoutput=1 

\usepackage[english]{babel}
\usepackage{jheppub}
\usepackage[T1]{fontenc}
\usepackage{graphicx,calc}
\usepackage{epsfig}
\usepackage{amsmath}
\usepackage{amssymb}
\usepackage{float}
\usepackage{placeins}
\usepackage{braket}
\usepackage{slashed}
\usepackage{mathdots}
\usepackage{lipsum}
\usepackage{feynmf}
\usepackage{cleveref}
\usepackage{bbm}
\usepackage{color}
\usepackage{multicol}
\usepackage{caption}
\usepackage{array}
\usepackage[export]{adjustbox}
\usepackage{subcaption}
\usepackage{multicol}
\usepackage{comment}
\usepackage{colortbl}
\usepackage[most]{tcolorbox}
\definecolor{myred}{RGB}{168,4,4}
\definecolor{myblue}{RGB}{48,53,149}

\usepackage{soul} 

\newcommand{\beq}{\begin{equation}}
\newcommand{\eeq}{\end{equation}}

\makeatletter
\newcommand\xleftrightarrow[2][]{%
  \ext@arrow 9999{\longleftrightarrowfill@}{#1}{#2}}
\newcommand\longleftrightarrowfill@{%
  \arrowfill@\leftarrow\relbar\rightarrow}
\makeatother

\allowdisplaybreaks

\title{Towards numerical two-loop integrand reduction}

\author[a]{Giuseppe Bevilacqua,}
\author[b]{Dhimiter Canko,}
\author[a]{Costas G. Papadopoulos,}
\author[a,c,d]{Aris Spourdalakis}
\affiliation[a]{Institute of Nuclear and Particle Physics, NCSR Demokritos, Patr. Grigoriou E' \& 27 Neapoleos Str, 15341 Agia Paraskevi, Greece}
\affiliation[b]{Dipartimento di Fisica e Astronomia, Università di Bologna e INFN, Sezione di Bologna, Via Irnerio 46, I-40126 Bologna, Italy}
\affiliation[c]{University of Debrecen, Faculty of Science and Technology, Department of Experimental Physics, 4010, Debrecen, PO Box 105, Hungary}
\affiliation[d]{Institute for Theoretical Physics, ELTE Eötvös Loránd University, Pázmány Péter sétány 1/A, H-1117 Budapest, Hungary}
\emailAdd{bevilacqua@inp.demokritos.gr}
\emailAdd{dhimiter.canko2@unibo.it}
\emailAdd{costas.papadopoulos@inp.demokritos.gr}
\emailAdd{aris.spourdalakis@ttk.elte.hu}

\abstract{We present a method for the integrand-level reduction of two-loop helicity amplitudes in both $d=4-2\epsilon$ and $d=4$ dimensions. The amplitude is expressed in terms of a set of Feynman integrals and their coefficients that depend on the external kinematics. The analysis presented in this paper, in conjunction with the ongoing development of the computational framework {\tt HELAC-2LOOP}, paves the road for the construction of an automated program for two-loop amplitude calculations for arbitrary scattering processes.}


\begin{document} 
\maketitle
\flushbottom

\section{Introduction}
\label{Introduction}

Several decades after its establishment, the Standard Model (SM) keeps offering a successful description of Strong and Electroweak interactions among fundamental particles. One of the most concrete and reliable ways of probing the SM is the comparison of its predictions to experimental data from high-energy particle collisions. The present energy frontier is represented by the Large Hadron Collider (LHC) with its four experiments, CMS, ATLAS, ALICE, and LHCb. The forthcoming High Luminosity upgrade of the LHC (HL-LHC) will significantly boost the statistical accuracy of the experimental data that are being collected, allowing experiments to reach percent-level precision in many cases~\cite{Dainese:2019rgk, Caola:2022ayt, CMS:2025hfp}. The accurate interpretation of the data from the LHC, as well as from possible future colliders~\cite{FCC:2025lpp}, demands theoretical predictions of comparable precision. 

Theoretical predictions are formulated from first principles within the framework of perturbative Quantum Field Theory, where scattering cross sections and other observables are computed as a power series expansion in the coupling constants of the SM. The first term in this expansion corresponds to the leading-order (LO) contribution, followed by subsequent corrections at next-to-leading order (NLO), next-to-next-to-leading order (NNLO), and so on. The calculation of NNLO corrections (and occasionally even N$^3$LO) is necessary in order to meet the current (and near future) experimental precision. The bottleneck for most NNLO calculations consists, at present, of the computation of double-virtual corrections, which rely upon two-loop dimensionally regularized scattering amplitudes. As the number of particles in a process increases, the complexity of the latter increases dramatically. The current frontier is the calculation of $2 \to 3$ processes~\cite{Huss:2025nlt}, with several results in this direction being achieved in recent years \cite{Badger:2017jhb, Badger:2018enw, Abreu:2018jgq, Abreu:2018zmy, Abreu:2019odu, Badger:2019djh, Hartanto:2019uvl, Chawdhry:2020for, Kallweit:2020gcp, Abreu:2021oya, Badger:2021nhg, Badger:2021imn, Badger:2021ega, Chawdhry:2021mkw, Chawdhry:2021hkp, Czakon:2021mjy, Abreu:2021asb, Badger:2022ncb, Badger:2023mgf, Abreu:2023bdp, Agarwal:2023suw, DeLaurentis:2023nss, DeLaurentis:2023izi, Badger:2024sqv, Badger:2024mir, Mazzitelli:2024ura, Badger:2024dxo, Agarwal:2024jyq, DeLaurentis:2025dxw} as the product of considerable effort from the theoretical community.

The calculation of two-loop amplitudes can be regarded as a three-step procedure. The initial step involves the construction of the amplitude through the use of standard Feynman graph generation~\cite{Hahn:2000kx, Nogueira:1991ex} or recursive approaches~\cite{Pozzorini:2022ohr, Canko:2023lvh}. The second step includes the reduction of the amplitude into a set of independent Feynman integrals, commonly known as \textit{master integrals}. For this to be achieved, the constructed amplitude should be projected into a set of scalar Feynman integrals, which can be done using tensor~\cite{Peraro:2019cjj, Peraro:2020sfm, Anastasiou:2023koq, Goode:2024mci, Goode:2024cfy}, or integrand~\cite{Kosower:2011ty, Mastrolia:2011pr, Zhang:2012ce, Badger:2012dp, Mastrolia:2012an, Mastrolia:2012wf, Kleiss:2012yv, Mastrolia:2013kca, Badger:2013gxa, Ita:2015tya, Mastrolia:2016dhn, Peraro:2019okx, Abreu_2021_C} reduction methods. The generated scalar integrals from these methods need to be furthermore reduced, at the integral level, into master integrals using  \textit{integration-by-parts} (IBP) identities \cite{Tkachov:1981wb, Chetyrkin:1981qh, Laporta:2001dd}, satisfied by the Feynman integrals within dimensional regularization ($d=4-2\epsilon$). The latter field has benefited from the application of \textit{finite-field techniques}~\cite{vonManteuffel:2014ixa, Peraro:2016wsq} for solving IBP identities, and the recent advancements in the generation of optimized IBP systems, utilizing \textit{syzygy equations}~\cite{Larsen:2015ped, Wu:2023upw} and the \textit{block-triangular form}~\cite{Liu:2018dmc, Guan:2024byi}. The last step for the computation of the scattering amplitude is the evaluation of the master integrals resulting from the reduction, which can be attained by methods such as \textit{sector decomposition}~\cite{Binoth:1999sp, Heinrich:2008si, Heinrich:2023til}, \textit{differential equations}~\cite{Barucchi:1973zm, Kotikov:1990kg, Kotikov:1991hm, Gehrmann:1999as, Henn:2013pwa, Moriello:2019yhu}, \textit{auxiliary mass flow}~\cite{Liu:2017jxz, Liu:2022chg}, and \textit{dimension-changing transformation}~\cite{Huang:2024qan, Huang:2024nij}.

The present work focuses on the second step of the aforementioned procedure, and more specifically on the integrand reduction of two-loop scattering amplitudes into scalar integrals. In the one-loop case, the integrand reduction \textit{OPP method}~\cite{Ossola:2006us}, demonstrated a systematic algebraic approach to the computation of amplitudes, based upon solving \textit{cut equations}, meaning equations resulting from setting some (or all) of the inverse propagators to zero, following a top-down approach. 
Integrand reduction has been implemented in various numerical algorithms~\cite{Ossola:2007ax,Ellis:2007br, Ellis:2008ir,Berger:2008sj,vanHameren:2009dr,vanHameren:2009vq, Mastrolia:2010nb, Badger:2010nx, Bevilacqua:2011xh, Peraro:2014cba, Alwall:2014hca, GoSam:2014iqq, Buccioni:2019sur} and has boosted  automation of NLO computations. Our goal is to extend the OPP method to two-loop calculations~\cite{Kosower:2011ty, Mastrolia:2011pr, Zhang:2012ce, Badger:2012dp, Mastrolia:2012an, Mastrolia:2012wf, Kleiss:2012yv, Mastrolia:2013kca, Badger:2013gxa, Ita:2015tya, Mastrolia:2016dhn}. We investigate two-loop amplitude reduction at the integrand level, in $d=4-2\epsilon$ and $d=4$ dimensions, by solving cut equations in a top-down approach~\cite{Kosower:2011ty, Mastrolia:2011pr, Zhang:2012ce, Badger:2012dp, Badger:2013gxa, Ita:2015tya, Mastrolia:2016dhn, Abreu_2021_C} in several two-loop examples.
In general, to facilitate computations, Feynman integrals that share the same topological and kinematical structure can be grouped into \textit{integral families}. A novel element of our study is the extension of the cut equations to include all inverse propagators characterizing a provided integral family and not just the ones of the Feynman graph under consideration. This approach yields a larger number of systems of equations to be solved, albeit smaller in size, and has the advantage of allowing the reduction of all the Feynman graphs that belong to the same family altogether. Furthermore, we study an alternative method for integrand-level reduction that does not rely on cut equations, but rather on sampling the integrand's numerator over random values of loop momenta.

This paper is structured as follows. In section 2, the notation employed is introduced, and the general ideas and concepts of the methods studied are explained. In section 3, the application of these methods to several examples with different kinematics is presented, ranging from four- to six-particle processes. In section 4, we conclude by giving a summary of our study and its prospects for the future.

\section{Description of the method}
\label{Method}
\subsection{Basic notation and conventions}
A generic unrenormalized $L$-loop scattering amplitude can be written schematically as
\beq
{\cal A}^{(L)}(\{p\})=\sum_{g=1}^{G} \, \left( \int \prod_{i=1}^L \, [dk_i] \, \frac{{\cal N}(k_1,\dots,k_L,\{p\})}{D_1\, D_2 \, \cdots D_n} \,\right)_g \, ,
\label{amplitude}
\eeq
where $k_1,\dots,k_L$ are loop momenta, $[dk_i]$ denotes the integration measure associated to the loop momentum $k_i$, and $\{p\}$ represents the set of external kinematics. The amplitude consists generally of $G$ partial contributions, individually denoted by $g$. Depending on the considered approach, the latter may be individual Feynman diagrams or consistently defined sub-amplitudes. In the following, we will refer to them as \textit{loop topologies}. In what follows, we focus on the genuine two-loop contributions to the amplitude (so-called Theta topologies), namely those that cannot be trivially reduced to the product of one-loop ones (Infinity and Dumbbell topologies)\footnote{For the terminology, see Figure 1 of Ref.~\cite{Bevilacqua:2024fec}.}. Each loop topology is characterized by a \textit{numerator} ${\cal N}$, which is a function of external momenta and wavefunctions as well as loop momenta, and by a product of $n$ \textit{inverse propagators} $D_1, \dots, D_n$, appearing in its denominator\footnote{Inverse propagators raised in higher integer powers must also be included to accommodate special contributions.}. 

The first step towards constructing a reduction method for two-loop calculations is to parametrize the numerator $\mathcal{N}$ in terms of the inverse propagators, moving along the path set by Ref.~\cite{Ossola:2006us}. A few comments are in order at this point. The numerator consists of independent scalar products involving loop and external momenta and possibly transverse vectors, i.e., $k_i \cdot k_j$, $k_i \cdot p_j, k_i \cdot \eta_j$. The number $N$ of independent scalar products among loop and external momenta is easily derived, depending on the considered regularization scheme. Denoting with $L$ the number of loops and with $E$ the number of external momenta, in the Conventional Dimensional Regularization (CDR) scheme
\beq
N=L(L+1)/2+L (E-1) \,,
\eeq
whereas in the t'Hooft-Veltman (tHV) scheme
\beq
N=
\begin{cases}
L(L+1)/2+L (E-1) & \text{for} \, \,  \, E<5 \\
L(L+1)/2+4 L & \text{for} \, \, \, E \ge 5
\end{cases} .
\eeq

In one-loop calculations ($L=1$), the number of inverse propagators $n$ appearing in the denominator of Eq.~(\ref{amplitude}) is always larger than or equal to $N$. Consequently, all scalar products appearing in the numerator ${\cal N}$, besides the so-called spurious terms~\cite{Ossola:2006us}, can be expressed in terms of the inverse propagators $D_i$, meaning they are all \textit{reducible scalar products} (RSP). Conversely, at higher loops ($L \ge 2$), part of the aforementioned scalar products are not expressible in terms of the $D_i$'s. The latter are commonly denoted \textit{irreducible scalar products} (ISP).

Extending the OPP approach to two loops, one can express the numerator $\mathcal{N}$ of a generic two-loop integrand in the following form:
\beq
{\cal N}=P^{(n)}+\sum_{i=1}^n P_i^{(n-1)}D_i+\sum_{i=1}^{n-1}\sum_{j>i}^n P_{ij}^{(n-2)}D_iD_j+\ldots +P_{12 \ldots n}^{(0)} D_1 D_2 \ldots D_n
\label{integrand reduction}
\eeq
The equation holds, as in the case of one loop~\cite{Ossola:2006us}, whenever the numerator of the amplitude has a rank, in $k_1$ and $k_2$, that does not exceed the number of corresponding propagators. In fact, as we will see later, the expansion may end long before its final contribution, depending on the rank of the numerator in question (see also Eq.~(\ref{eq:multi-cut}) and section~\ref{Applications}).  
Plugging this expression in Eq.~(\ref{amplitude}) results in a splitting of the integrand into terms with a different number of denominators.  
The quantities $P^{(n)}$, $P_i^{(n-1)}$, $P_{ij}^{(n-2)},\dots$ appearing in Eq.~(\ref{integrand reduction}) are \textit{polynomials} of the form
\begin{equation}
P = \sum_{l=1}^{M} \, b_l \, m_l, \,    
\label{eq:monomials}
\end{equation}
where $b_i = b_i(\{p\})$ are coefficients which depend on the external kinematics, and the monomials $m_i$ are built upon the ISP characteristic of each term,
and possibly of scalar products of the form $k_i\cdot \eta_j$.

Focusing on the two-loop case, our goal is to decompose the amplitude, ${\cal A}^{(2)}$, in terms of a set of Feynman integrals $F_i$ and coefficients $c_i$ that depend only on the external kinematics,
\beq
{\cal A}^{(2)}=\sum_i c_i(\{p\}) \, F_i \,,
\label{integral reduction}
\eeq
where the $F_i$'s take the form 
\beq
F_i \equiv F(a_1,\ldots,a_N)=\int \prod_{i=1}^2 [dk_i]\frac{1}{D_1^{a_1}\ldots D_N^{a_N}},\;\;\;\;a_i\in \mathbb{Z} \,.
\label{integral}
\eeq
This can be achieved by appropriately expressing the monomials $m_l$ in Eq.~(\ref{eq:monomials}) in terms of the inverse propagators $D_i$~\cite{Sotnikov:2019onv}. Eq.~(\ref{integral reduction}) is at the core of \textit{numerical} methods for two-loop computations. Provided that the integrals $F_i$ are known (or a procedure to reduce the latter to a subset of master integrals is established), determining the coefficients $c_i$ at \textit{integrand level} helps to address the problem of two-loop computations in a general, process-independent way. This idea expands upon well-established methods developed for one-loop calculations, such as OPP reduction \cite{Ossola:2006us}.

In the following subsections, we will introduce two different methods that can be used to determine the polynomials $P$ appearing in Eq.~(\ref{integrand reduction}) and to fit their coefficients. 

\subsection{Linear fit and fit by cut approach}
\label{fit by cut}
We first address the question of determining the polynomials $P$ and fitting their coefficients in Eq.~(\ref{integrand reduction}), assuming that the numerator is known \textit{analytically}. This allows us to construct the polynomials $P$ directly from the symbolic expressions appearing in the numerator. Moving along the same path of one-loop reduction, the fitting is carried out by iteratively applying cut equations, namely setting denominators to zero. Starting with the so-called maximal cut equation,
\beq
D_1=D_2=\ldots=D_n=0 \,,
\label{cuteqn}
\eeq
we can identify the coefficients of $P^{(n)}$,
\beq
P^{(n)}={\cal N}|_{D_1=D_2=\ldots=D_n=0} \,.
\label{maxcut}
\eeq
Then we can iteratively fit the rest of the polynomials by appropriately subtracting the terms computed in the previous step. For instance, the first next-to-maximal contribution reads  
\beq
P^{(n-1)}_1=\left. \left(\frac{{\cal N}-{\cal N}|_{D_1=D_2=\ldots=D_n=0}}{D_1}\right)\right\rvert_{D_2=\ldots=D_n=0} \,,
\label{nextmaxcut}
\eeq
and so on. We note that cut equations, Eq.~(\ref{cuteqn}), translate into a system of linear relations among scalar products $k_i\cdot k_j$ and $k_i\cdot p_j$, where $k_i$ and $p_j$ denote generically loop and external momenta, respectively. This allows us to straightforwardly solve Eq.~(\ref{maxcut}), Eq.~(\ref{nextmaxcut}), and the rest of the equations resulting from all the sub-maximal cuts, by substitution rules, for any process. We call this procedure of solving Eq.~(\ref{integrand reduction}), a {\it linear fit}. We will show in the following sections how this applies to the case of 4-, 5-, and 6-particle scattering amplitudes.  

In certain instances, the analytical calculation of the numerator ${\cal N}$ can prove to be a highly challenging task. In such cases, resorting to a \textit{numerical} computation, facilitated by dedicated software packages such as  {\tt HELAC-2LOOP} \cite{Canko:2023lvh}, can be a feasible alternative. In this case, there are two issues to be addressed:
\begin{enumerate}
\item the solutions of cut equations must be expressed in a form suitable for numerical evaluation of numerators;
\item  the polynomials $P$ appearing in  Eq.~(\ref{integrand reduction}) must be constructed without \textit{a priori} analytical knowledge of the numerator.
\end{enumerate}
To address the first issue, we need a suitable representation of the loop momenta. Given two arbitrary massless momenta $l_1^\mu,l_2^\mu$, let us  define
\begin{equation}
\begin{gathered}
l_3^\mu=\bar{u}_-(l_1)\gamma^\mu u_-(l_2) \,,
\\
l_4^\mu=\bar{u}_-(l_2)\gamma^\mu u_-(l_1) \,.
\end{gathered}
\label{eq:l3l4}
\end{equation}
The set $\{l_1^\mu, l_2^\mu, l_3^\mu, l_4^\mu\}$ forms a basis in $d=4$ dimensions. This allows us to express the loop momenta $k_1,k_2$ as follows,
\begin{equation}
\begin{gathered}
    k_1=x_1 \, l^{(1)}_1+x_2 \, l^{(1)}_2+x_3 \, l^{(1)}_3+x_4 \, l^{(1)}_4 \\
    k_2=y_1 \, l^{(2)}_1+y_2 \, l^{(2)}_2+y_3 \, l^{(2)}_3+y_4 \, l^{(2)}_4
\end{gathered}
\label{eq:basis}
\end{equation}
where the coefficients $x_i$ and $y_i$ are expressible in terms of scalar products of the form $k_i\cdot p_j$. The latter coefficients characterize the loop momenta in $d=4$ dimensions. Complemented by $\mu_{11}$, $\mu_{12}$ and $\mu_{22}$\footnote{In $d=4-2\epsilon$, $\mu_{ij}\equiv k^{(\epsilon)}_i\cdot k^{(\epsilon)}_j $, with $k^{(\epsilon)}$ the components of the loop momenta beyond $d=4$ dimensions.}, they form a set of eleven variables which characterizes completely the loop momenta in $d=4-2\epsilon$ dimensions: $\vec{X}=\{ x_1,x_2,x_3,x_4,y_1,y_2,y_3,y_4,\mu_{11},\mu_{12},\mu_{22}\}$. 
We solve cut equations\footnote{For alternative approaches, see Refs.~\cite{Kosower:2011ty, Mastrolia:2011pr, Zhang:2012ce, Badger:2012dp, Badger:2013gxa, Ita:2015tya}.} in terms of these variables, as we will see in \cref{Applications}. In $d=4-2\epsilon$, the solution to the cut equations is unique in terms of the ISP, whereas in $d=4$ we usually have disjoint branches~\cite{Zhang:2012ce}, see \cref{Double-box} for an explicit example.

The parametrization of the polynomials $P$ in terms of the ISP is obtained through the program {\tt BasisDet}~\cite{Zhang:2012ce}. The latter provides a set of monomials which take the form $\prod_i x_i^{r_i}$, where $x_i$ denote ISP and $r_i$ is an integer ranging from zero to some upper value calculated from the maximal tensor rank of the polynomial $P$ with respect to $k_1$, $k_2$ and $k_1,k_2$ combined\footnote{Corresponding to {\tt RenormalizationCondition} in ref. \cite{Zhang:2012ce}.}. For example, a term such as $(k_1\cdot p_1)^2 (k_2\cdot p_3)^3$ has a tensor rank 2 for $k_1$, 3 for $k_2$ and 5 for $k_1,k_2$ combined.

At one loop, $P$ consists of terms depending solely on the external kinematics and the so-called spurious terms, which are specific to each cut. The spurious terms, although necessary for the reduction at the integrand level, do not contribute to the final result as they integrate to zero. The final result is determined by the coefficients that depend only on the external kinematics and multiply the appropriately chosen basis of integrals. At two loops, the existence of spurious terms that integrate to zero is less straightforward: there are certainly spurious terms compiled by the loop momenta and the transverse directions over the external momenta, whenever present. Nevertheless, the simple one-loop picture is spoiled by the fact that the integrals in Eq.~(\ref{integral}) obey a set of IBP identities, resulting in a set of master integrals, which are then evaluated using different techniques. In this paper, we do not address the issue of IBP reduction: we assume that all the integrals resulting from the integrand reduction are, or will be, available numerically, as happens for the one-loop case.  

Returning to the solution of Eq.~(\ref{integrand reduction}), let us first address the case of $d=4-2\epsilon$ dimensions. The cut equations fix a subset of the eleven parameters needed to fully describe the loop momenta.
Assuming that the set of monomials $m_i$ $(i=1, \ldots, M)$ parametrizing a given polynomial $P$ is established, then an $M\times M$ matrix, $\cal M$, is obtained by evaluating the monomials on the solution to the cut equation, 
by assigning $M$ random values to the free parameters of the vector $\vec{X}$, 
obtaining thus $M$ instances of it, i.e. $\vec{X}_j,\; j=1,\ldots,M$, and then computing the elements of the matrix ${\cal M}$, as follows: ${\cal M}_{i,j}\equiv m_i(\vec{X}_j)$. The numerator ${\cal N}(\vec{X},d)$ can be cast in the form
\begin{equation}
\label{numerator split}
{\cal N}\equiv {\cal N}(\vec{X},d)={\cal N}_{0}+\sum_{i\ge1}\epsilon^i {\cal N}^{(i)}_{\epsilon} \, ,
\end{equation}
by expanding in powers of $\epsilon\equiv (4-d)/2$, where both ${\cal N}_{0}\equiv {\cal N}(\vec{X},d)|_{d=4}$ and ${\cal N}^{(i)}_{\epsilon}\equiv {\cal N}^{(i)}_{\epsilon}(\vec{X})$, are accessible numerically and depending on $\mu_{ij}$ through $\vec{X}$. These terms are used to calculate the $M\times 1$ matrices, ${\cal B}^{(0)}_j ={\cal N}_{0}(\vec{X}_j)$, ${\cal B}^{(i)}_j ={\cal N}^{(i)}_{\epsilon}(\vec{X}_j)$. Then the given polynomial $P$ is written explicitly as 
\beq
P=\sum_{i=1}^M\left(c_i^{(0)}+\sum\epsilon^j \;c_i^{({j})}\right) m_i
\eeq
where 
\beq
{\vec c\;}^{(0)}={\cal M}^{-1}{\cal B}^{(0)}\;\;\;\;\;
{\vec c\;}^{({i})}={\cal M}^{-1}{\cal B}^{(i)}\;\;\;\;\;
\eeq
After the whole iterative procedure is completed, the so-called $N=N$ test is performed. The latter consists of checking the validity of Eq.~(\ref{integrand reduction}) for arbitrary assignment of numerical values for all free parameters of the loop kinematics, $\vec{X}$, not restricted by cut equations. We have checked that the $N=N$ test is fulfilled when the polynomials $P$ are constructed directly from the analytic expression of the numerator, as well as when using {\tt BasisDet} to construct the ansatz for the polynomials.

In $d=4$ we obtain several disjoint solutions of the cut equations in terms of $\vec{X}^{(d=4)}=\{ x_1,x_2,x_3,x_4,y_1,y_2,y_3,y_4\}$. On each branch, we have checked analytically that the $d=4$ numerator, ${\cal N}_{4,0}\equiv {\cal N}(\vec{X},d)|_{\mu_{ij}=0,d=4}$, assumes a different form. On the other hand, the set of monomials obtained previously in $d=4-2\epsilon$ dimensions, contains linear dependencies due to the fact that in $d=4$ the two loop momenta and the three independent external momenta, in a 4-particle amplitude for instance, is an over complete set and Gram determinants among them vanish, leading to non-trivial relations.
In that case, the set of monomials $m_i, \; i=1,\ldots,M$, provided by {\tt BasisDet}, is evaluated at each branch of the cut-equation solution. Assuming the existence of $r$ branches, the matrix $\cal M$ of size $(r M)\times M$ and the matrix ${\cal B}_0^{(4)}$ of size  $(r M)\times 1$, are calculated using ${\cal M}_{i,j}\equiv m_i(\vec{X}^{(d=4)}_j)$, ${\cal B}_{0j}^{(4)}\equiv {\cal N}_{4,0}(\vec{X}^{(d=4)}_j)$, with $i=1,\ldots,M,\;j=1,\ldots,rM$. 
The system  
\beq
{\cal M} \, {\vec c\;}^{(d=4)}={\cal B}_0^{(4)}
\eeq
can still be solved with standard Linear Algebra algorithms such as QR decomposition, as long as the rank of the matrix is full, namely $rank({\cal M})=M$. As we will see later, this is true in most cases, but solutions can still be obtained in cases where the matrix is rank-deficient, $rank({\cal M})<M$. Notice that in $d=4-2\epsilon$ case, when the information on the dependence of the numerator on $d$ and $\mu_{ij}$ is available, the reduction of the amplitude is complete, whereas in four dimensions, where this information is not available, the so-called rational terms need to be calculated in addition~\cite{Ossola:2008xq, Badger:2008cm, Pozzorini:2020hkx, Lang:2020nnl, Lang:2021hnw}. 
\subsection{Global fit approach}
\label{global fit}
Assuming that the numerator ${\cal N}$ is a function of the eleven (eight) parameters in $d=4-2\epsilon$ ($d=4$) dimensions, and a numerator provider (such as {\tt HELAC-2LOOP}) is capable of returning numerical values for ${\cal N}_4$ and ${\cal N}_\epsilon$, we can instead determine all the coefficients of a single polynomial $P$ by means of a \textit{global fit}~\cite{Peraro:2019okx, Badger:2021imn}. In this method, all monomials involved in the polynomial $P$ are evaluated for several arbitrary random numerical assignments of the above-mentioned parameters. In this way, we construct a single square matrix ${\cal M}$, with size $M\times M$, where $M$ is the number of the monomials appearing in $P$. Using the same random values of the parameters, we also sample the numerator functions ${\cal N}_4$, ${\cal N}_\epsilon$, ${\cal N}_{4,0}$ and build $M\times 1$ matrices that we denote ${\cal B}^{(4)}$, ${\cal B}^{(\epsilon)}$ and ${\cal B}_0^{(4)}$ respectively.
The difference to the fit by cut approach is that the kinematics are not constrained to obey cut equations. While the number $M$ of monomials is much larger, since it accommodates all monomials from all polynomials in Eq.~(\ref{integrand reduction}), there are no significant conceptual differences between the two approaches. As we will show later, the solution can be obtained by using standard Linear Algebra packages, such as {\tt Eigen}~\cite{eigenweb} and {\tt LAPACK}~\cite{LaPack}, at both double and quadruple precision.
\subsection{Projecting over a full family}
\label{projecting}
From the perspective of the one-loop OPP approach, Eq.~(\ref{integrand reduction}) addresses the reduction of the numerator in Eq.~(\ref{amplitude}) in terms of the $n$ inverse propagators $D_i$ appearing in it. As we already pointed out, the one-loop case is special in the sense that the number of independent scalar products $N$ and the number of inverse propagators $n$ obey the relation $N \le n$. Thus, all scalar products can be expressed in terms of the $D_i$'s, which appear in the denominator of the loop integrand. Starting from two loops, $N>n$, and thus one is left with a set of ISP that cannot be expressed as above. However, one can define an enlarged set of inverse propagators such that \textit{all} scalar products are expressible as combinations of the latter. This enlarged set of inverse propagators is named \textit{family}.
We can consider projecting the numerator over the full family of inverse propagators:
\beq
{\cal N}=P^{(N)}+\sum_{i=1}^N P_i^{(N-1)}D_i+\sum_{i=1}^{N-1}\sum_{j>i}^N P_{ij}^{(N-2)}D_iD_j+\ldots +P_{12\ldots N}^{(0)} D_1 D_2 \ldots D_N
\label{integrand reduction over family}
\eeq
For example, cutting over 7 propagators in a two-loop four-particle topology,
\beq
\begin{gathered}
{\cal N}_{4p}=P_{7c}(v_1,v_2,u_1,u_2)+\sum_{i=1}^{7} P_{6c}^{(i)}(v_1,v_2,u^i_1,u^i_2,u^i_3)D_i
\\+\sum_{i<j}^{7} P_{5c}^{(i,j)}(v_1,v_2,u^{i,j}_1,u^{i,j}_2,u^{i,j}_3,u^{i,j}_4)D_iD_j
\\+\sum_{i<j<k}^{7} P_{4c}^{(i,j,k)}(v_1,v_2,u^{i,j,k}_1,\ldots,u^{i,j,k}_5)D_iD_jD_k
\\
+\sum_{i<j<k<l}^{7} P_{3c}^{(i,j,k,l)}(v_1,v_2,u^{i,j,k,l}_1,\ldots,u^{i,j,k,l}_6)D_iD_jD_kD_l
\\+\sum_{i<j<k<l<m}^{7} P_{2c}^{(i,j,k,l,m)}(v_1,v_2,u^{i,j,k,l,m}_1,\ldots,u^{i,j,k,l,m}_7)D_iD_jD_kD_lD_m+\ldots
\end{gathered} 
\label{eq:4p7p}
\eeq
with $v_i\equiv k_i\cdot\eta$, where $\eta$ is the vector transverse to $p_1,p_2,p_3$ necessary to complete the 4-dimensional physical space, while $u^{i}_I$, $u^{i,j}_I$, $u^{i,j,k}_I$, $u^{i,j,k,l}_I, \, \dots$,  stand for ISP in the corresponding cut,
will generically lead to polynomials $P$ depending on more variables rather than cutting over all 9 propagators,
\beq
\begin{gathered}
{\cal N}_{4p}=P_{9c}(v_1,v_2)+\sum_{i=1}^{9} P_{8c}^{(i)}(v_1,v_2,u_i)D_i
+\sum_{i<j}^{9} P_{7c}^{(i,j)}(v_1,v_2,u_i,u_j)D_iD_j
\\+\sum_{i<j<k}^{9} P_{6c}^{(i,j,k)}(v_1,v_2,u_i,u_j,u_k)D_iD_jD_k
\\+\sum_{i<j<k<l}^{9} P_{5c}^{(i,j,k,l)}(v_1,v_2,u_i,u_j,u_k,u_l)D_iD_jD_kD_l
\\
+\sum_{i<j<k<l<m}^{9} P_{4c}^{(i,j,k,l,m)}(v_1,v_2,u_i,u_j,u_k,u_l,u_m)D_iD_jD_kD_lD_m+\ldots
\end{gathered}
\label{eq:4p9p}
\eeq
where now $u_i$ stand for ISP in the corresponding cut (see explicit expressions in section~\ref{Applications}).

The same is true for a five-particle amplitude, where now the physical phase-space is fully described by the four external independent momenta, cutting over 8 propagators results in
\beq
\begin{gathered}
{\cal N}_5=P_{8c}(u_1,u_2,u_3)+\sum_{i=1}^{8} P_{7c}^{(i)}(u^i_1,\ldots,u^i_4)D_i
\\+\sum_{i<j}^{8} P_{6c}^{(i,j)}(u^{i,j}_1,\ldots,u^{i,j}_5)D_iD_j
\\+\sum_{i<j<k}^{8} P_{5c}^{(i,j,k)}(u^{i,j,k}_1,\ldots,u^{i,j,k}_6)D_iD_jD_k
\\
+\sum_{i<j<k<l}^{8} P_{4c}^{(i,j,k,l)}(u^{i,j,k,l}_1,\ldots,u^{i,j,k,l}_7)D_iD_jD_kD_l
\\
+\sum_{i<j<k<l<m}^{8} P_{3c}^{(i,j,k,l,m)}(u^{i,j,k,l,m}_1,\ldots,u^{i,j,k,l,m}_8)D_iD_jD_kD_lD_m+\ldots
\end{gathered}
\label{eq:5p8p}
\eeq
while cutting over all 11 propagators, which defines the 5-point family, results in
\beq
\begin{gathered}
{\cal N}_5=P_{11c}+\sum_{i=1}^{11} P_{10c}^{(i)}(u_i)D_i+\sum_{i<j}^{11} P_{9c}^{(i,j)}(u_i,u_j)D_iD_j
\\+\sum_{i<j<k}^{11} P_{8c}^{(i,j,k)}(u_i,u_j,u_k)D_iD_jD_k
\\
+\sum_{i<j<k<l}^{11} P_{7c}^{(i,j,k,l)}(u_i,u_j,u_k,u_l)D_iD_jD_kD_l
\\
+\sum_{i<j<k<l<m}^{11} P_{6c}^{(i,j,k,l,m)}(u_i,u_j,u_k,u_l,u_m)D_iD_jD_kD_lD_m +\ldots
\end{gathered}
\label{eq:5p11p}
\eeq
Notice that in this case $P_{11c}$ depends only on the external kinematics, since the maximal cut determines all degrees of freedom of the loop momenta. 
We observe that projecting over the full family has the effect of simplifying the structure of the individual polynomials $P$, since they depend on a reduced number of ISP.
Moreover, all the ISP $u_i$ in Eq.~(\ref{eq:5p11p}), can be expressed in terms of the inverse propagators of the family, leading directly to a variation of the master equation,   
\beq
{\cal N}_5=c^{(0)}+\sum_{i=1}^{11}\sum_{a}c^{(i,a)} D^{a}_{i}+\sum_{i<j}^{11}\sum_{a,b} c^{(i,j,a,b)}D_i^{a}D_j^{b}+\ldots
\label{eq:multi-cut}
\eeq
where, first the coefficients $c$ depend only on the external kinematics and second the inverse Feynman propagators, $D_i$, enter not linearly as before, but with powers determined by the fit-by-cut procedure.
The amplitude is then given by,
\beq
{\cal A}^{(2)}_5=c^{(0)}F[a_1,\ldots,a_{11}]+\sum_{i=1}^{11}\sum_{a} c^{(i,a)}F[a_1,\ldots,a_i-a,\ldots,a_{11}]+\ldots
\label{eq:amplitude}
\eeq
with $a_1=\ldots=a_8=1$ and $a_9=a_{10}=a_{11}=0$ and $F$ given as in Eq.(\ref{integral}). Eqs.~(\ref{eq:5p8p}-\ref{eq:amplitude}) are straightforwardly extensible to any $n-$particle amplitude with $n>5$, see section~\ref{6-gluon} for details in case $n=6$. 
\section{Application of the method}
\label{Applications}

In this section, the application of the method described in \cref{Method} is demonstrated through several examples. 
The considered numerators are representative of a variety of scattering processes and consist of different kinematic dependencies, ranging from four-point to six-point kinematics. For the generation of the analytic expressions required, we used the \texttt{Mathematica} packages \texttt{FeynArts}~\cite{Hahn:2000kx}  and \texttt{FeynCalc}~\cite{Shtabovenko:2023idz}, except for the 6-particle case, section \ref{6-gluon}, which has been generated by \texttt{FORM}~\cite{Ruijl:2017dtg}. 

\subsection{Four-Point Kinematics}
\label{Four_point}
\subsubsection{Double-box topologies}
\label{Double-box}

\begin{figure}[t!]
\centering
\begin{subfigure}[b]{0.328\textwidth}
\centering
\includegraphics[scale=0.36]{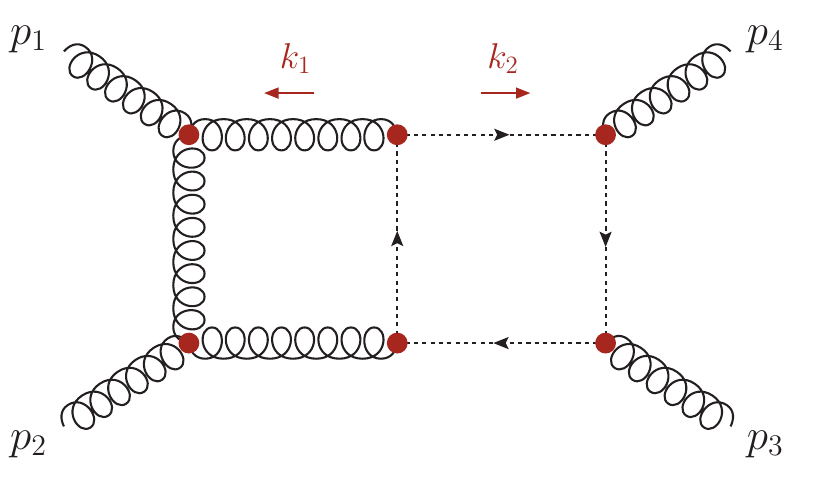}
\end{subfigure}
\hfill
\begin{subfigure}[b]{0.328\textwidth}
\centering
\includegraphics[scale=0.36]{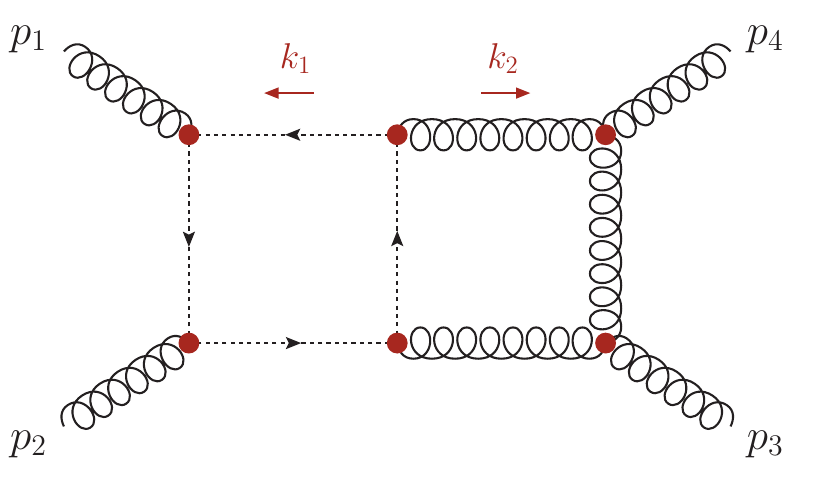}
\end{subfigure}
\hfill
\begin{subfigure}[b]{0.328\textwidth}
\centering
\includegraphics[scale=0.36]{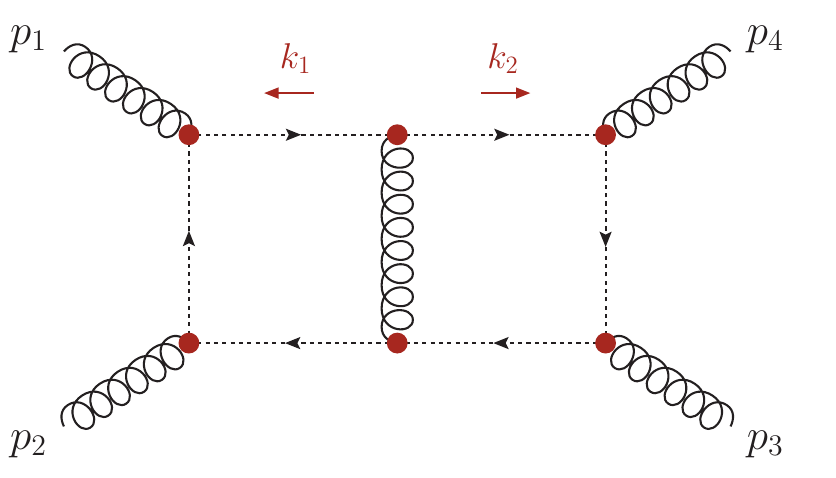}
\end{subfigure}
\begin{subfigure}[b]{0.328\textwidth}
\centering
\includegraphics[scale=0.36]{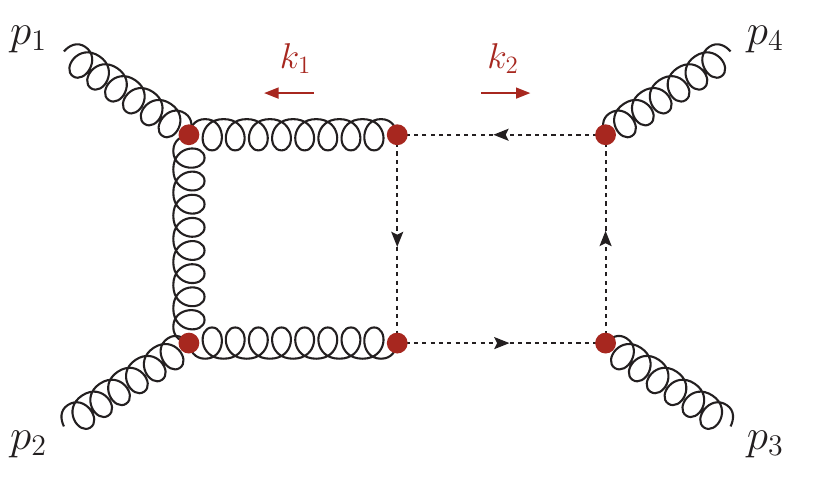}
\end{subfigure}
\hfill
\begin{subfigure}[b]{0.328\textwidth}
\centering
\includegraphics[scale=0.36]{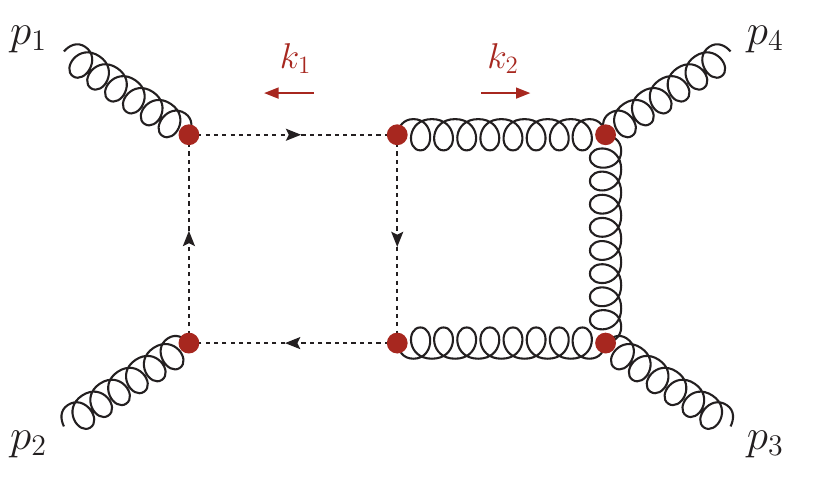}
\end{subfigure}
\hfill
\begin{subfigure}[b]{0.328\textwidth}
\centering
\includegraphics[scale=0.36]{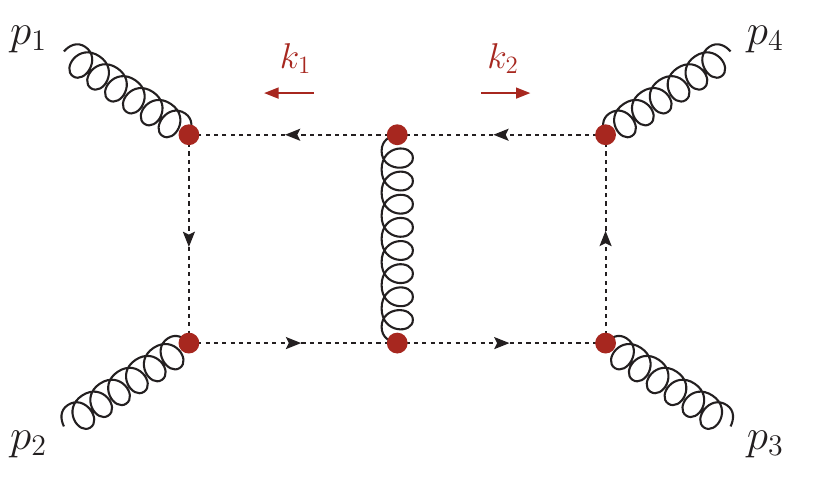}
\end{subfigure}
\hfill
\begin{subfigure}[b]{0.328\textwidth}
\centering
\includegraphics[scale=0.36]{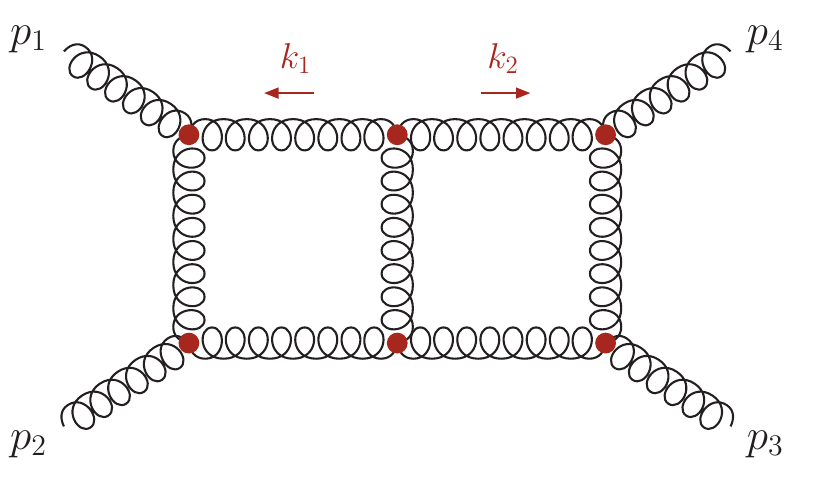}
\end{subfigure}
\caption{Feynman graphs contributing to the double-box numerator under study. There are seven contributions, considering gluons and (anti-)ghosts running within the loops. Curly lines denote gluons, while dotted lines denote ghosts (arrow aligned with the momentum flow) and anti-ghosts (arrow in the opposite direction of the momentum flow).}
\label{fig:doublebox}
\end{figure}

Here we focus on the numerator of the double-box topology constructed by the seven Feynman graphs depicted in Fig.~\ref{fig:doublebox}. This numerator contributes to the scattering amplitude of the process $gg \to gg$. The inverse propagators describing the family for this topology can be chosen as
    \begin{equation}
        \begin{gathered}
            D_1=k_1^2, \quad D_2=\left(k_1+p_1\right){}^2, \quad D_3=\left(k_1+p_{12}\right){}^2, \quad D_4=\left(k_1+k_2\right){}^2, \quad D_5=k_2^2,\\            
            D_6=\left(k_2-p_{123}\right){}^2, \quad D_7=\left(k_2-p_{12}\right){}^2, \quad D_8=\left(k_2-p_1\right){}^2, \quad D_9=\left(k_1+p_{123}\right){}^2
        \end{gathered} \,
    \label{eq:4p-planar-family}
    \end{equation}
Above and henceforth, the shorthand notation $p_{i \ldots j}=p_i + \ldots + p_j$ will be used to denote the sum of the incoming external on-shell momenta, and $s=(p_1+p_2)^2$ and $t=(p_2+p_3)^2$ the standard Mandelstam variables. For all analyses presented below we have used the following helicity assignment: $\lambda_1=+1,\lambda_2=+1,\lambda_3=-1,\lambda_4=-1$, for the helicities of the external gluons, all assumed to be incoming.

\begin{enumerate}
    \item Linear fit

    We seek to solve Eq.~(\ref{eq:4p7p}) using cut equations and the known analytic expression of the numerator. 
 
    The maximal cut equations
    \begin{equation}
        D_1=D_2=D_3=D_4=D_5=D_6=D_7=0
    \end{equation}
    result in determining seven invariants
    \begin{equation}
    \begin{gathered}    
        k_1\cdot k_1\to 0, \quad k_1\cdot k_2\to 0, \quad k_1\cdot p_1\to 0, \quad k_1\cdot p_2\to -\frac{s}{2}, \quad k_2\cdot k_2\to 0,\\ k_2\cdot p_2\to \frac{s}{2}-k_2\cdot p_1, \quad k_2\cdot p_3\to -\frac{s}{2}.
        \label{eq:7cut}
        \end{gathered}
    \end{equation}
    By applying the above relation on both sides of Eq.~(\ref{eq:4p7p}), we can fully determine the polynomial $P_7$.
    The latter consists of 70 coefficients over the ISP monomials $\{k_1\cdot p_3, \; k_1\cdot \eta, \; k_2\cdot p_1, \; k_2\cdot \eta\}$. 

    Subtracting $P_7$ in Eq.~(\ref{eq:4p7p}), we can now determine the polynomials of kind $P_6$ in the same way. There are seven six-cuts and therefore seven $P_6$ polynomials to determine. As an example, the first six-cut,
    \begin{equation}
        D_2=D_3=D_4=D_5=D_6=D_7=0
    \end{equation}
    leads to
    \begin{equation}
        \begin{gathered}
       k_2\cdot p_2\to \frac{s}{2}-k_2\cdot p_1, \quad k_1\cdot p_2\to -\frac{s}{2}, \quad k_2\cdot p_3\to -\frac{s}{2},\\ k_1\cdot k_2\to k_1\cdot p_1, \quad k_1\cdot k_1\to -2 k_1\cdot p_1, \quad k_2\cdot k_2\to 0 \,,
    \end{gathered}
\end{equation}
    where there are now 5 ISP: $\{k_1\cdot p_2, \; k_1\cdot p_3, \; k_1\cdot \eta , \; k_2\cdot p_1, \; k_2\cdot \eta\}$. The polynomial $P^{(6)}_1$ consists of 111 coefficients.
    This process is iterated until the level of a two-cut, after which all resulting polynomials vanish.
    The data for all cuts are summarized in Tab.~\ref{tab:double-box-data}. The analytic solution for the polynomials satisfies explicitly Eq.~(\ref{eq:4p7p}).
\begin{table}[t!]
    \centering
    \begin{tabular}{|c|c|c|c|}
    \hline
        Level & Number of cuts & Number of coefficients & Scaling \\ \hline
         7 & 1 & 70 & 4,4,4 \\
         6 & 7  & 695 & 3,3,4\\
         5 & 21 & 1430 & 3,3,3\\
         4 & 35 & 1017 & 2,2,2\\
         3 & 35 & 225 & 1,1,1\\
         2 & 21 & 9 & 0,0,0\\
         \hline
    \end{tabular}
    \caption{Double-box linear fit information beginning with 7-cut. The numbers in the last column refer to the maximum powers of $k_1$, $k_2$, and $k_1, k_2$ combined, as described in the text. }
    \label{tab:double-box-data}
\end{table}

We now seek to solve Eq.~(\ref{eq:4p9p}), namely projecting over all 9 propagators in the double-box family. This has the advantage of building a reduction procedure that covers all 4-particle planar diagrams, including the double-box, the penta-triangle and the hexa-bubble. The maximal cut equations read 
\begin{equation}
        D_1=D_2=D_3=D_4=D_5=D_6=D_7=D_8=D_9=0
    \end{equation}
which leads to 
\begin{equation}
    \begin{gathered}    
        k_1\cdot k_1\to 0, \quad k_1\cdot k_2\to 0, \quad k_1\cdot p_1\to 0, \quad k_1\cdot p_2\to -\frac{s}{2}, \quad k_1\cdot p_3\to \frac{s}{2},\\ k_2\cdot k_2\to 0, \quad k_2\cdot p_1\to 0, \quad k_2\cdot p_2\to \frac{s}{2}, \quad k_2\cdot p_3\to -\frac{s}{2}
        \label{eq:9cut}
        \end{gathered}
    \end{equation}
    
 By applying the above relations on both sides of Eq.~(\ref{eq:4p9p}), we can fully determine the polynomial $P_9$. The latter consists of 13 coefficients over the ISP monomials $\{k_1\cdot \eta, \; k_2\cdot \eta\}$. Subtracting as before $P_9$ in Eq.~(\ref{eq:4p9p}), we can now determine the polynomials $P_8$ in the same way. There are nine 8-cuts and therefore nine $P_8$ polynomials to determine. As an example, the first 8-cut,
     \begin{equation}
        D_2=D_3=D_4=D_5=D_6=D_7=D_8=D_9=0
    \end{equation}
    leads to
    \begin{equation}
        \begin{gathered}
       k_2\cdot p_1\to 0, \quad k_2\cdot p_2\to \frac{s}{2}, \quad k_1\cdot p_2\to -\frac{s}{2}, \quad k_2\cdot p_3\to -\frac{s}{2}, \quad k_1\cdot p_3\to \frac{s}{2},\\ k_1\cdot k_2\to k_1\cdot p_1, \quad k_1\cdot k_1\to -2 k_1\cdot p_1, \quad k_2\cdot k_2\to 0 \,,
    \end{gathered}
\end{equation}
where there are now 3 ISP: $\{k_1\cdot p_1, \; k_1\cdot \eta, \; k_2\cdot \eta\}$. 

\begin{table}[t!]
    \centering
    \begin{tabular}{|c|c|c|c|}
    \hline
        Level & Number of cuts & Number of coefficients & Scaling \\ \hline
         9 & 1 & 13 & 4,4,4 \\
         8 & 9  & 227 & 4,4,4\\
         7 & 36 & 963 & 3,3,3\\
         6 & 84 & 1445 & 2,2,2\\
         5 & 126 & 780 & 1,1,1\\
         4 & 126 & 116 & 0,0,0\\
         \hline
    \end{tabular}
    \caption{Double-box linear fit information beginning with 9-cut.}
    \label{tab:double-box-data-9}
\end{table}

The data for all cuts are summarized in Tab.~\ref{tab:double-box-data-9}. The analytic solutions for the polynomials satisfy explicitly Eq.~(\ref{eq:4p7p}).
The total number of non-zero coefficients is slightly larger than previously, namely 3544 versus 3446.  We have verified that after reducing by IBP identities the integrals appearing in Eqs.~(\ref{eq:4p7p},\ref{eq:4p9p}), the coefficients of the top-sector master integrals coincide with those obtained from {\tt Caravel}~\cite{Abreu_2021_C}, for all helicity assignments~\footnote{In {\tt Caravel} the results are given for the color-stripped helicity amplitude, whereas in our case we have studied a subset of the contributions to the amplitude, as given in Fig.~\ref{fig:doublebox}. Nevertheless, a comparison of the top-sector master integral coefficients is possible since all other Feynman graphs do not contribute to them.}   
     
    \item Fit by cut in $d=4-2\epsilon$ dimensions

Let us now assume that the numerator is available only using a numerical approach, as the one implemented in {\tt HELAC-2LOOP}, including terms proportional to $\mu_{ij}$ and $\epsilon=(d-4)/2$. Then the realization of the solutions of the cut equations, Eq.~(\ref{eq:7cut}) in a numerical setup is based on the determination of the four-dimensional part of the loop momenta following Eq.~(\ref{eq:basis}). In fact the solution for any cut has a unique analytic form in terms of ISP. The 7 cut, Eq.~(\ref{eq:7cut}), reads as follows:
\begin{equation}
    \begin{gathered}
        k_1\cdot p_1\to 0, \quad k_2\cdot p_2\to \frac{1}{2} \left(s-2 k_2\cdot p_1\right), \quad k_1\cdot p_2\to -\frac{s}{2}, \quad k_2\cdot p_3\to -\frac{s}{2},\\ \mu _{11}\to -\frac{\frac{4 s \left(k_1\cdot p_3\right){}^2}{s+t}-4 s k_1\cdot p_3+4 t \left(k_1\cdot \eta \right){}^2+s (s+t)}{4 t},\\
        \mu _{12}\to \frac{\frac{k_1\cdot p_3 \left(4 (s+2 t) k_2\cdot p_1-2 s t\right)}{s+t}+t \left(s-4 k_1\cdot \eta  \, k_2\cdot \eta \right)-2 s k_2\cdot p_1}{4 t},\\ \mu _{22}\to -\frac{4 s t k_2\cdot p_1+4 s \left(k_2\cdot p_1\right){}^2+t \left(4 (s+t) \left(k_2\cdot \eta \right){}^2+s t\right)}{4 t (s+t)} 
    \end{gathered}
\end{equation}
For comparison, the cut conditions for 9 propagators, Eq.~(\ref{eq:9cut}),  are 
\begin{equation}
    \begin{gathered}
        k_2\cdot p_1\to 0, \quad k_1\cdot p_1\to 0, \quad k_2\cdot p_2\to \frac{s}{2}, \quad k_1\cdot p_2\to -\frac{s}{2}, \\ 
        k_2\cdot p_3\to -\frac{s}{2}, \quad k_1\cdot p_3\to \frac{s}{2}, \quad \mu _{11}\to -\left(k_1\cdot \eta \right){}^2-\frac{s t}{4 (s+t)}, \\  \quad \mu _{12}\to \frac{s t}{4 (s+t)}-k_1\cdot \eta  k_2\cdot \eta, \quad  \mu _{22}\to -\left(k_2\cdot \eta \right){}^2-\frac{s t}{4 (s+t)}
    \end{gathered}
\end{equation}
In terms of the basis introduced in section~\ref{fit by cut}, \textit{i.e.} $\{x_1,\dots,x_4,y_1,\dots,y_4,\mu_{11},\mu_{12},\mu_{22}\}$, the solution takes the form 
\begin{equation}
    \begin{gathered}
       x_1\to -1, \quad x_2\to 0, \quad y_1\to 0, \quad y_2\to -1, \quad \mu _{11}\to 4 s x_3 x_4, \quad \mu _{22}\to 4 s y_3 y_4,
       \\ \mu _{12}\to (x_3+x_4-y_3-y_4) r -2 (s+t) (x_4 y_4 + x_3 y_3) - 2t\left(x_3 y_4+x_4 y_3\right)-t/2
    \end{gathered}
\end{equation}
with $r=\sqrt{-t (s+t)}$. In a numerical setup, for instance, calculating on the kinematic point $s=1,\, t=-1/5$, the 7-cut is represented by 
    \begin{equation}
    \begin{gathered}
    x_1\to -1, \quad x_2\to 0, \quad y_1\to 0, \quad y_2\to -1, \quad \mu _{11}\to 4 x_3 x_4, \quad \mu _{22}\to 4 y_3 y_4,\\ \mu _{12}\to \left(4 x_3 \left(-4 y_3+y_4+1\right)+4 x_4 \left(y_3-4 y_4+1\right)-4 y_3-4 y_4+1\right)/10
    \end{gathered}
  \end{equation}
We can now determine the coefficients of the polynomial $P_7$, which in this case has 70 coefficients, by calculating the numerator ${\cal N}$, and the monomials $m_i$ of the basis, using 70 random assignments of the undetermined variables, $x_3,x_4,y_3,y_4$, \textit{i.e.} ${\cal N}_i$ and $M_{ij}$ respectively, and solving the corresponding matrix equation 
\begin{equation}
    \sum_{j=1}^{70}M_{ij} \, c_j={\cal N}_i,\;\;\;i=1,\ldots,70
\end{equation}
for the unknown coefficients $c$. The full-rank matrix $M$ is straightforwardly invertible, and the solution checked agrees to the numerical precision used against the analytic result. We have confirmed that this way we can calculate all the coefficients of Tab.~\ref{tab:double-box-data} numerically. The same is true for the case of projecting over the 9 propagators, see Tab.~{\ref{tab:double-box-data-9}}. 
    \item Fit by cut in $d=4$ dimensions
\begin{table}[t!]
    \centering
    \begin{tabular}{cc}
        Level & RenormalizationCondition \\
         7 & \{\{\{1,0\},4\},\{\{0,1\},4\},\{\{1,1\},4\}\} \\
         6 & \{\{\{1,0\},4\},\{\{0,1\},4\},\{\{1,1\},4\}\} \\
         5 & \{\{\{1,0\},3\},\{\{0,1\},3\},\{\{1,1\},3\}\}\\
         4 & \{\{\{1,0\},2\},\{\{0,1\},2\},\{\{1,1\},2\}\}\\
         3 & \{\{\{1,0\},1\},\{\{0,1\},1\},\{\{1,1\},1\}\} \\
         2 & \{\{\{1,0\},0\},\{\{0,1\},0\},\{\{1,1\},0\}\} \\
    \end{tabular}
    \caption{Input used for {\tt BasisDet}, denoting by $\{\{1,0\},n_1\}$ the maximal power $n_1$ in $k_1$, by $\{\{0,1\},n_2\}$ the maximal power $n_2$ in $k_2$ and by $\{\{1,1\},n_{12}\}$ the maximal power $n_{12}$ in $k_1$, $k_2$ combined.}
    \label{tab:rencond}
\end{table}  
    
In this case, the monomials that make up the polynomials of the analytic solution described either by the linear fit or the fit by cut in $d=4-2\epsilon$, are not independent, due to the existence of Gram-determinant identities which hold in $d=4$ dimensions. This is why for $d=4$ we use {\tt BasisDet} as a means to parametrize the polynomials, therefore no longer relying on analytical knowledge of the numerator in order to create the anzatz. By construction, this requires us to specify the highest powers of the loop momenta to the {\tt BasisDet} program. Following the data of Tab.~\ref{tab:double-box-data}, the so-called RenormalizationCondition in {\tt BasisDet} is fixed as shown in Tab.~\ref{tab:rencond}. Contrary to the $d=4-2\epsilon$ case discussed previously, in $d=4$  the cut equations do not admit a unique solution in terms of ISP. Using the same values for $s,t$ as above, the 7-cut solution is split into 6 branches,
\begin{equation*}
    \begin{gathered}
        \left\{x_1\to -1,x_2\to 0,x_3\to 0,x_4\to \frac{4 y_3-1}{4 \left(y_3+1\right)},y_1\to 0,y_2\to -1,y_4\to 0,y_3\to z(1)\right\},\\ \left\{x_1\to -1,x_2\to 0,x_3\to \frac{4 y_4-1}{4 \left(y_4+1\right)},x_4\to 0,y_1\to 0,y_2\to -1,y_3\to 0,y_4\to z(1)\right\},\\
        \left\{x_1\to -1,x_2\to 0,x_3\to 0,x_4\to -\frac{1}{4},y_1\to 0,y_2\to -1,y_3\to 0,y_4\to z(1)\right\}, \\ \left\{x_1\to -1,x_2\to 0,x_3\to -\frac{1}{4},x_4\to 0,y_1\to 0,y_2\to -1,y_4\to 0,y_3\to z(1)\right\},\\ 
 \left\{x_1\to -1,x_2\to 0,x_3\to 0,y_1\to 0,y_2\to -1,y_3\to 0,y_4\to \frac{1}{4},x_4\to z(1)\right\},\\ \left\{x_1\to -1,x_2\to 0,x_4\to 0,y_1\to 0,y_2\to -1,y_3\to \frac{1}{4},y_4\to 0,x_3\to z(1)\right\},
    \end{gathered}
\end{equation*}

each one parametrized by one undetermined variable $z(1)$. The polynomial $P_7$ is parametrized by only 28 coefficients, in $d=4$. By assigning 28 random values to $z(1)$ for each solution, we may form a matrix equation as follows:
\begin{equation}
    \sum_{j=1}^{28}M_{ij} \, c_j={\cal N}_i,\;\;\;i=1,\ldots,168
    \label{eq:dblfmatrixd4}
\end{equation}
where the rows of the matrix $M$ consist of the values of the monomials involved in the basis and $\cal N$ are the corresponding values of the numerator. Although this is a rectangular system of equations, it admits a unique solution for the coefficients $c$. We have checked against the analytic solution and found agreement up to the numerical precision used. The same is true for all lower-level cuts. 

    \item Global fit in $d=4-2\epsilon$ dimensions

As an alternative to determining the structure of Eq.~(\ref{eq:4p7p}), through cut equations, we have also studied the global fit, as described in the previous section.By expressing both sides of Eq.~(\ref{eq:4p7p}) in terms of the variables 
$V=\{x_1,x_2,x_3,x_4,y_1,y_2,y_3,y_4,\mu _{11},$ $\mu _{12},\mu _{22}\}$, parameterizing the loop momenta, we can form a matrix equation over all undetermined parameters $c$ entering in the definition of the polynomials $P_7,\ldots,P_2$ in  Eq.~(\ref{eq:4p7p}). The number of parameters $c$, as well as the monomials involved, is 3446. Assigning random values to all 11 variables of $V$, we form the equation 
\begin{equation}
    \sum_{j=1}^{3446}M_{ij} \, c_j={\cal N}_i,\;\;\;i=1,\ldots,3446
\end{equation}
where, as before, $M$ is the matrix formed by the numerical values of the monomials and $\cal N$ the numerical values of the numerator. The matrix $M$ is of full rank and the solution has been determined both in {\tt Mathematica} as well as using purely numerical tools, such as {\tt Eigen} and {\tt LAPACK}. In fact, the latter offers us the possibility to obtain our results in both double and quadruple precision. As a check of the validity of the solution, the $N=N$ test is successfully performed, in full precision, for arbitrary random assignment of the variables in $V$.   

    \item Global fit in $d=4$ dimensions

    A global fit in $d=4$ can be performed following the same line of reasoning as before. 
The total number of coefficients $c$ and monomials is 3018. 
The variables describing the loop momenta are $V=\left\{x_1,x_2,x_3,x_4,y_1,y_2,y_3,y_4\right\}$. Assigning random values to all 8 variables of $V$, we form the equation 
\begin{equation}
    \sum_{j=1}^{3018}M_{ij}c_j={\cal N}_i,\;\;\;i=1,\ldots,3018
    \label{eq:dbgbmatrixd4}
\end{equation}
where as before $M$ is the matrix formed by the numerical values of the monomials and $\cal N$ the numerical values of the numerator in $d=4$ dimensions. The matrix $M$ is of full rank and the solution has been determined both in {\tt Mathematica} as well as using purely numerical engines, such us {\tt Eigen} and {\tt LAPACK}. In fact the numerical engines offer us the possibility to obtain our results in both double and quadruple precision. As a check of the validity of the solution, the $N=N$ test is successfully performed, in full precision, for arbitrary random assignment of the variables in $V$.   
    
\end{enumerate}

\subsubsection{Penta-triangle topologies}
\label{Penta-triangle}

\begin{figure}[t!]
\centering
\begin{subfigure}[b]{0.328\textwidth}
\centering
\includegraphics[scale=0.38]{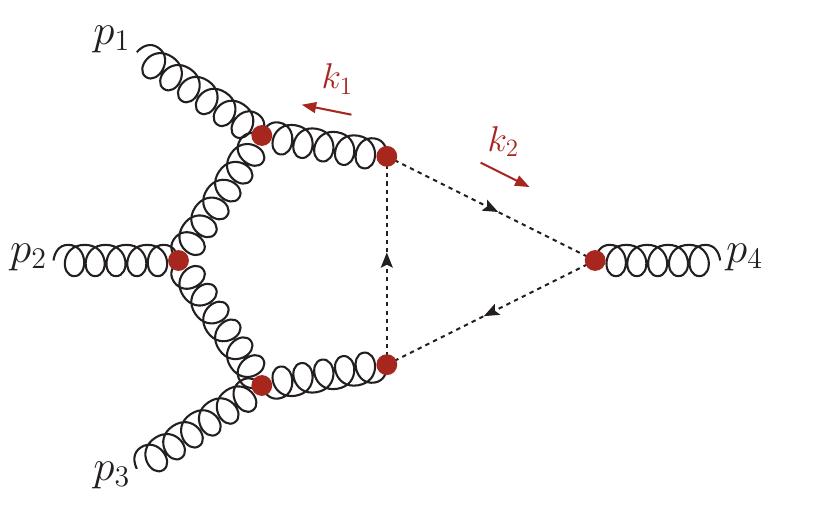}
\end{subfigure}
\hfill
\begin{subfigure}[b]{0.328\textwidth}
\centering
\includegraphics[scale=0.38]{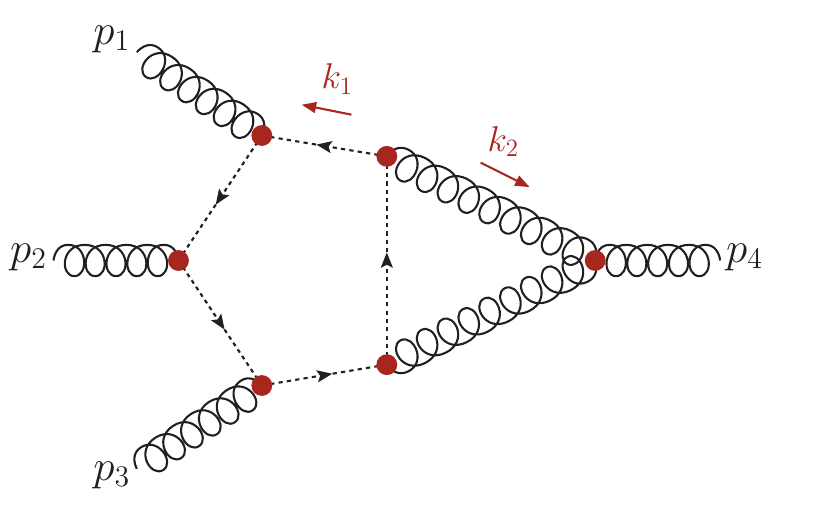}
\end{subfigure}
\hfill
\begin{subfigure}[b]{0.328\textwidth}
\centering
\includegraphics[scale=0.38]{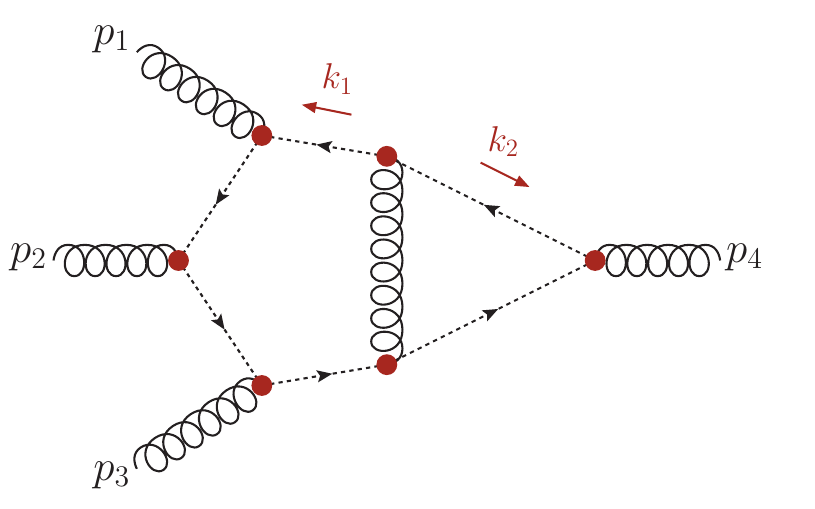}
\end{subfigure}
\begin{subfigure}[b]{0.328\textwidth}
\centering
\includegraphics[scale=0.38]{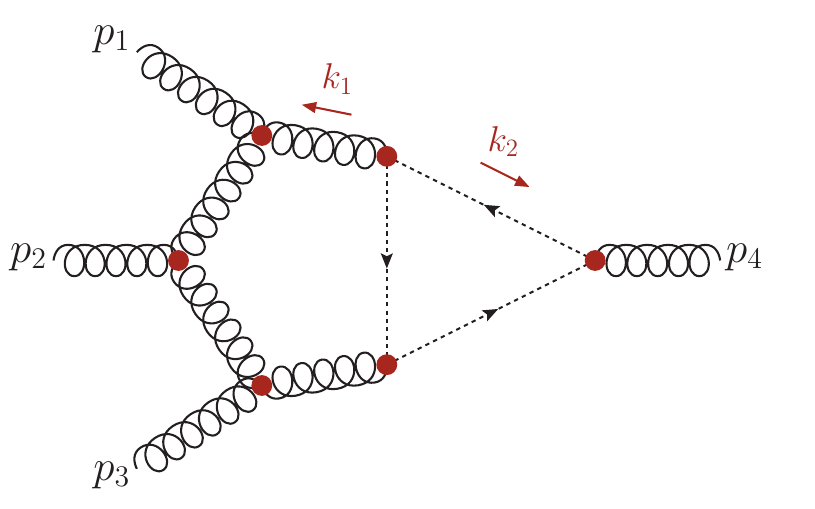}
\end{subfigure}
\hfill
\begin{subfigure}[b]{0.328\textwidth}
\centering
\includegraphics[scale=0.38]{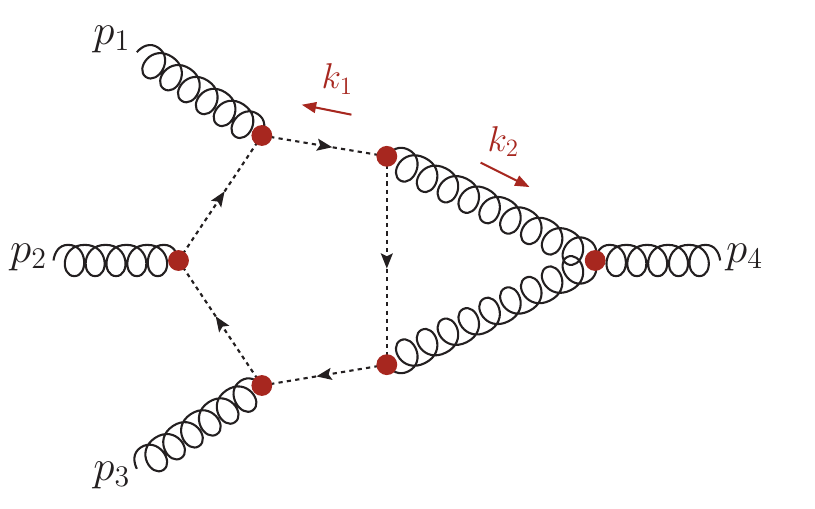}
\end{subfigure}
\hfill
\begin{subfigure}[b]{0.328\textwidth}
\centering
\includegraphics[scale=0.38]{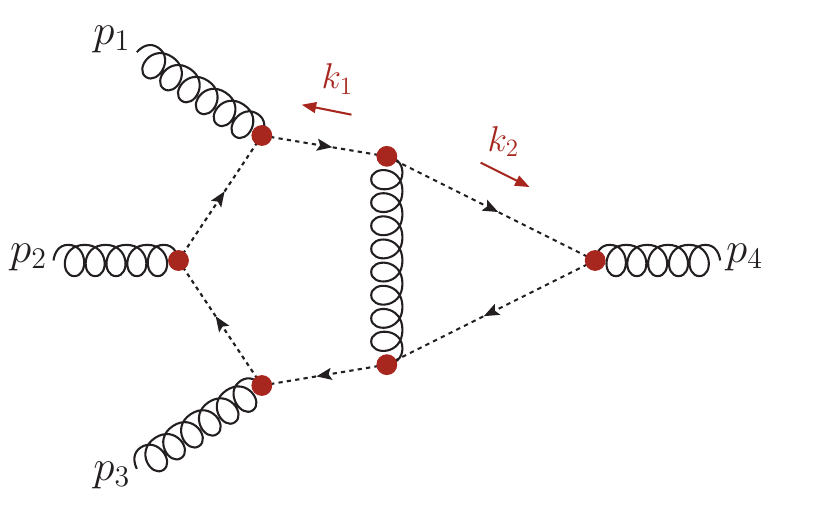}
\end{subfigure}
\hfill
\begin{subfigure}[b]{0.328\textwidth}
\centering
\includegraphics[scale=0.38]{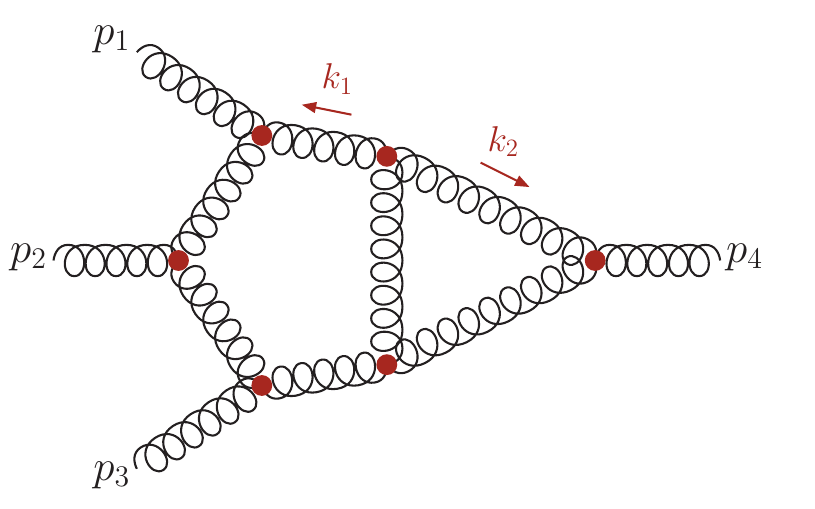}
\end{subfigure}
\caption{Feynman graphs contributing to the penta-triangle numerator under study. There are seven contributions, considering gluons and (anti-)ghosts running within the loops.}
\label{fig:pentatriangle}
\end{figure}

In this subsection, we study the numerator of the penta-triangle topology produced by the seven Feynman graphs depicted in Fig.~\ref{fig:pentatriangle}. This numerator is part of the scattering amplitude of the process $gg \to gg$. The helicity assignment is as in the case of double-box. The propagators describing the family for this topology coincide with the ones of Eq.~(\ref{eq:4p-planar-family}), but we choose a different ordering such that the auxiliary propagators appear in the end
\begin{equation*}
    \begin{gathered}
            D_1=k_1^2, \quad D_2=\left(k_1+p_1\right){}^2, \quad D_2=\left(k_1+p_{12}\right){}^2, \quad D_4=\left(k_1+p_{123}\right){}^2, \quad  D_5=\left(k_1+k_2\right){}^2, \\ D_6=k_2^2, \quad D_7=\left(k_2-p_{123}\right){}^2, \quad D_8=\left(k_2-p_{12}\right){}^2, \quad D_9=\left(k_2-p_1\right){}^2
    \end{gathered}
\end{equation*}
      
\begin{enumerate}
    \item Linear fit, fit by cut, and global fit in $d=4-2\epsilon$ dimensions
    
The $d=4-2\epsilon$ analysis is producing the same qualitative results as in the double-box case. For instance, the 7-cut is given by 
\begin{equation}
    \begin{gathered}
        k_1\cdot k_1\to 0, \quad k_1\cdot k_2\to 0, \quad k_1\cdot p_1\to 0, \quad k_1\cdot p_2\to -\frac{s}{2}, \quad k_1\cdot p_3\to \frac{s}{2},\\ k_2\cdot k_2\to 0, \quad k_2\cdot p_3\to -\left(k_2\cdot p_1\right)-k_2\cdot p_2
    \end{gathered}
\end{equation}
 The data regarding the number of cuts, coefficients of polynomials, and respective powers are collected in Tab.~\ref{tab:penta-triangle-data}.
 \begin{table}[t!]
    \centering
    \begin{tabular}{|c|c|c|c|}
    \hline
        Level & Number of cuts & Number of coefficients & Scaling \\ \hline
         7 & 1 & 53 & 4,4,4 \\
         6 & 7  & 573 & 4,4,4\\
         5 & 21 & 1249 & 3,3,3\\
         4 & 35 & 929 & 2,2,2\\
         3 & 35 & 206 & 1,1,1\\
         2 & 21 & 8 & 0,0,0\\
         \hline
    \end{tabular}
    \caption{Penta-triangle linear fit information beginning with 7-cut. }
    \label{tab:penta-triangle-data}
\end{table}

The corresponding data when projecting over 9 propagators are shown in Tab.~\ref{tab:penta-triangle-data9}.
 \begin{table}[t!]
    \centering
    \begin{tabular}{|c|c|c|c|}
    \hline
        Level & Number of cuts & Number of coefficients & Scaling \\ \hline
         7 & 1 & 9 & 3,3,3 \\
         6 & 9  & 573 & 4,3,4\\
         5 & 36 & 899 & 3,3,3\\
         4 & 84 & 1279 & 2,2,2\\
         3 & 126 & 652 & 1,1,1\\
         2 & 126 & 100 & 0,0,0\\
         \hline
    \end{tabular}
    \caption{Penta-triangle linear fit information beginning with 9 cut.}
    \label{tab:penta-triangle-data9}
\end{table}

    \item Fit by cut and global fit $d=4$

    In $d=4$, as before, we have to rely on {\tt BasisDet} in order to construct monomial bases. In this case, the numerator evaluated on all 7-cut solutions is zero. In the 6-cut sector, a new issue arises, namely, there are cuts for which the corresponding matrix equation, see Eq.~(\ref{eq:dblfmatrixd4}), is rank deficient, meaning that the solution is described in terms of undetermined coefficients. These undetermined coefficients, however, do not impact the $N=N$ test. The same is true for the global fit, see Eq.~(\ref{eq:dbgbmatrixd4}), where again the equation is rank deficient, nevertheless, its solution correctly describes the numerator. We note that the penta-triangle is the only case studied in this paper where such a problem occurs.

    This presents quite an interesting theoretical challenge for the construction of the ansatz in $d=4$: on the one hand, we have verified that choosing a loose ansatz, meaning a polynomial that leads to a rank deficient matrix, is in fact not a problem, as long as the system is solvable and the $N=N$ test is not affected. 
    On the other hand, restricting oneself to a basis of polynomials containing a smaller number of monomials, can lead to the following problem: while the solution to the linear equation may even be unique (i.e. full rank of the matrix ${\cal M}$), the functional behavior of the numerator may not be properly accounted for, leading to an incorrect solution which does not reproduce the $N=N$ test. The issue, therefore, of choosing an ansatz, i.e., a proper basis of polynomials to describe a given numerator is not at all trivial and still somewhat remains an open question, especially important if one aims to perform a $d=4$ fit. While {\tt BasisDet} seems to be the best option available, it still runs the risk of falling into one of the two above problems, depending on the inputs given. 
\end{enumerate}

\subsubsection{Hexa-bubble topologies}
\label{Hexa-bubble}

\begin{figure}[t!]
\centering
\begin{subfigure}[b]{0.24\textwidth}
\centering
\includegraphics[scale=0.36]{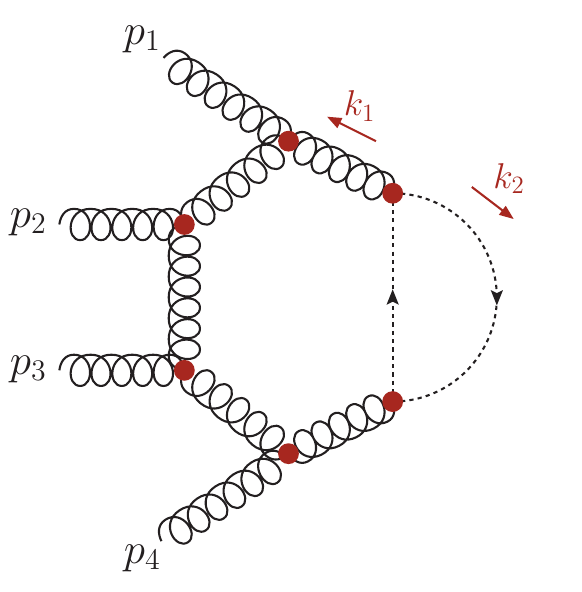}
\end{subfigure}
\hfill
\begin{subfigure}[b]{0.24\textwidth}
\centering
\includegraphics[scale=0.36]{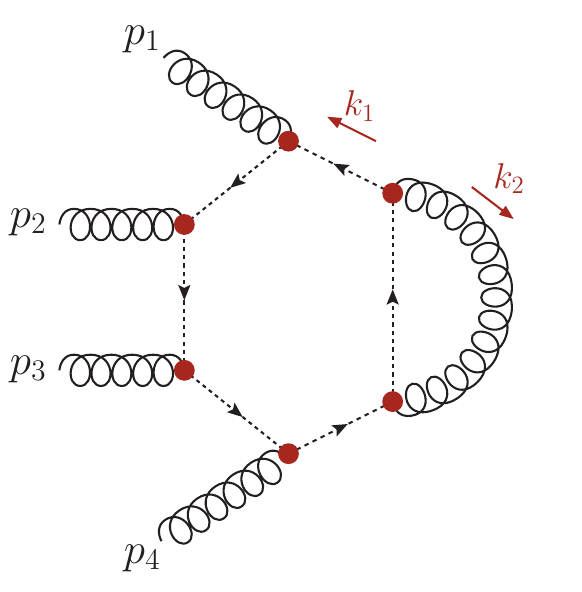}
\end{subfigure}
\hfill
\begin{subfigure}[b]{0.24\textwidth}
\centering
\includegraphics[scale=0.36]{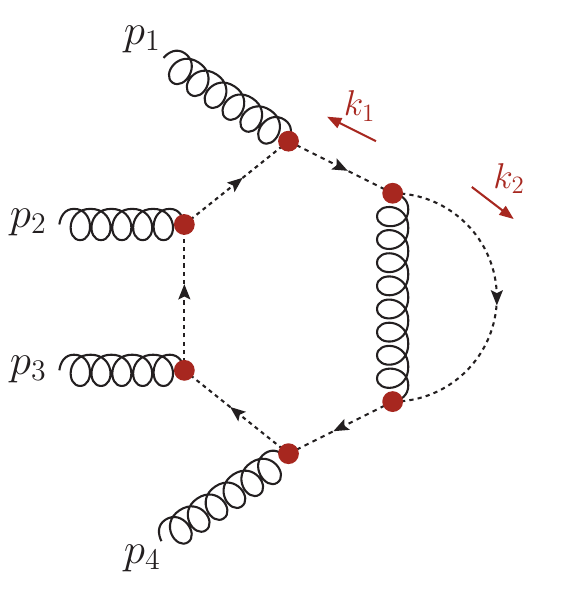}
\end{subfigure}
\hfill
\begin{subfigure}[b]{0.24\textwidth}
\centering
\includegraphics[scale=0.36]{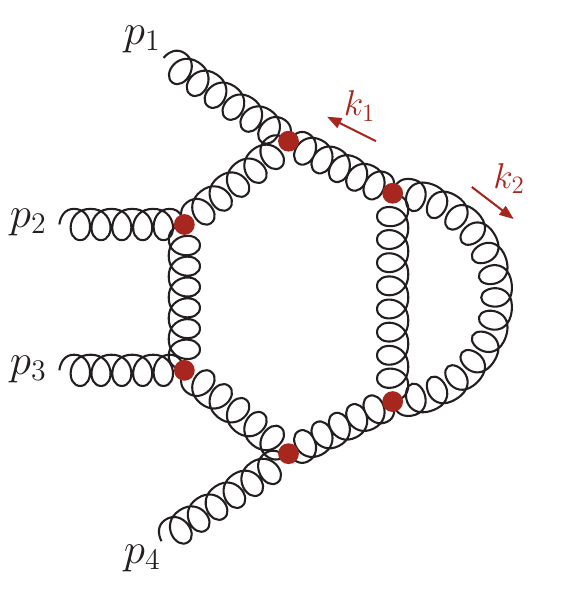}
\end{subfigure}
\caption{Feynman graphs contributing to the hexa-bubble numerator under study. There are four contributions, considering gluons and (anti-)ghosts running within the loops.}
\label{fig:hexabubble}
\end{figure}

We now consider the numerator of the hexa-bubble topology consisting of the four Feynman graphs depicted in Fig.~\ref{fig:hexabubble}. This numerator is part of the scattering amplitude of the process $gg \to gg$. The helicity assignment is as in the case of double-box. The inverse propagators describing this topology coincide with the ones of penta-triangle topology, but only the first six of those propagators appear in the denominator, with the first one being repeated and thus being raised to the second power. The analysis, projecting over 6 or 9 propagators, works for all cases, as in the double-box, in both $d=4-2\epsilon$ and $d=4$ dimensions. Of course, in the final expression of Eq.~(\ref{integral reduction}), integrals with a doubled propagator occur, which in any case have to be evaluated in all cases by IBP identities, or otherwise, independently from the integrand-level reduction discussed in this paper.  

\subsubsection{Non-planar double box topologies}
\label{Non-planar double box}

\begin{figure}[t!]
\centering
\begin{subfigure}[b]{0.328\textwidth}
\centering
\includegraphics[scale=0.37]{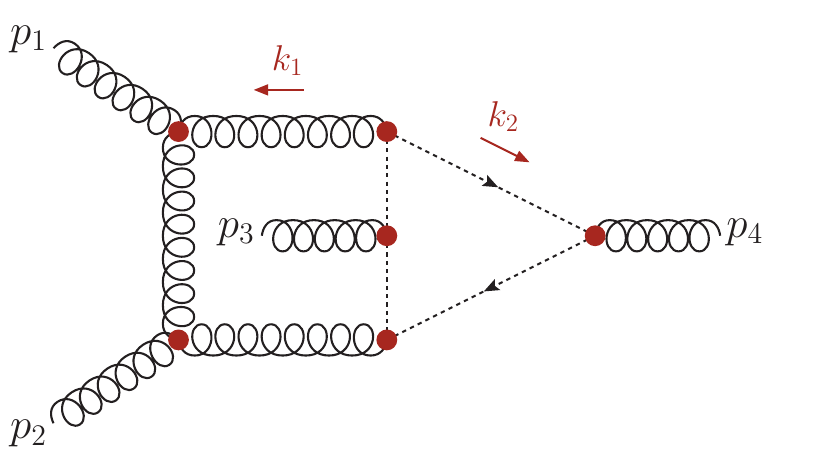}
\end{subfigure}
\hfill
\begin{subfigure}[b]{0.328\textwidth}
\centering
\includegraphics[scale=0.37]{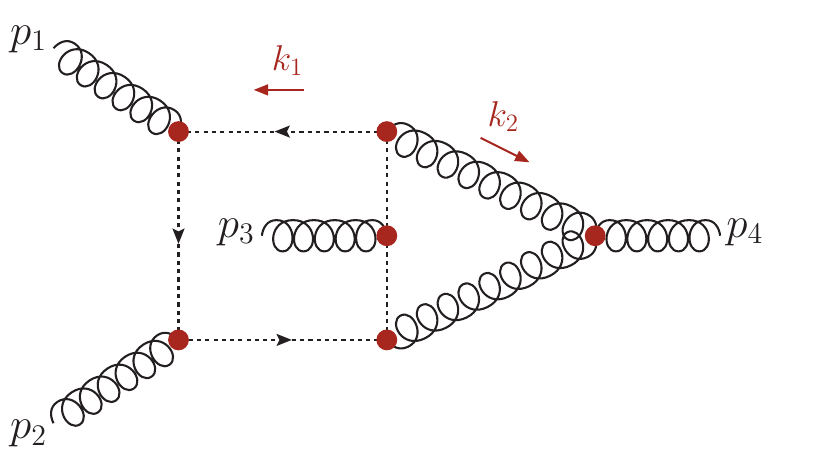}
\end{subfigure}
\hfill
\begin{subfigure}[b]{0.328\textwidth}
\centering
\includegraphics[scale=0.37]{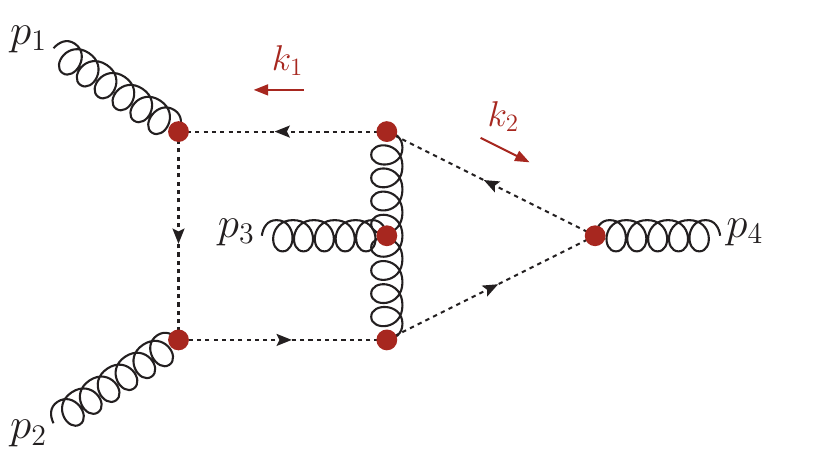}
\end{subfigure}
\begin{subfigure}[b]{0.328\textwidth}
\centering
\includegraphics[scale=0.37]{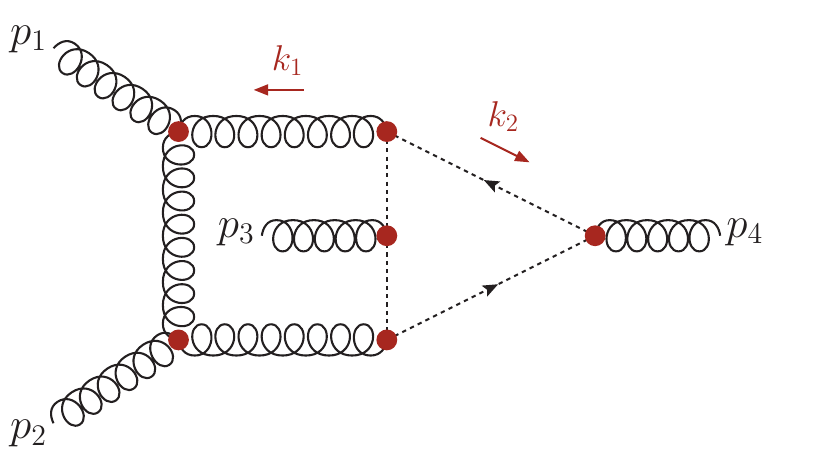}
\end{subfigure}
\hfill
\begin{subfigure}[b]{0.328\textwidth}
\centering
\includegraphics[scale=0.37]{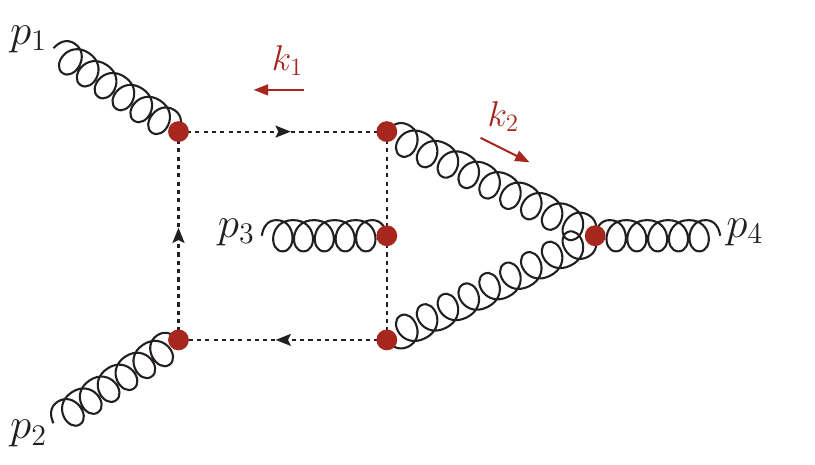}
\end{subfigure}
\hfill
\begin{subfigure}[b]{0.328\textwidth}
\centering
\includegraphics[scale=0.37]{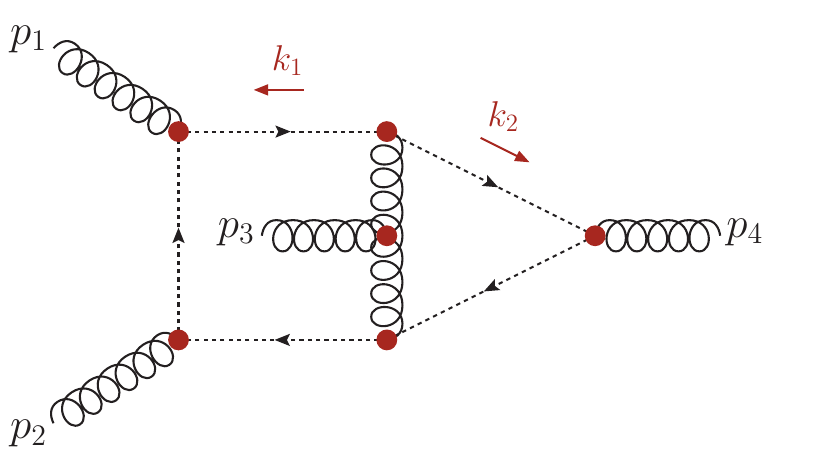}
\end{subfigure}
\hfill
\begin{subfigure}[b]{0.328\textwidth}
\centering
\includegraphics[scale=0.37]{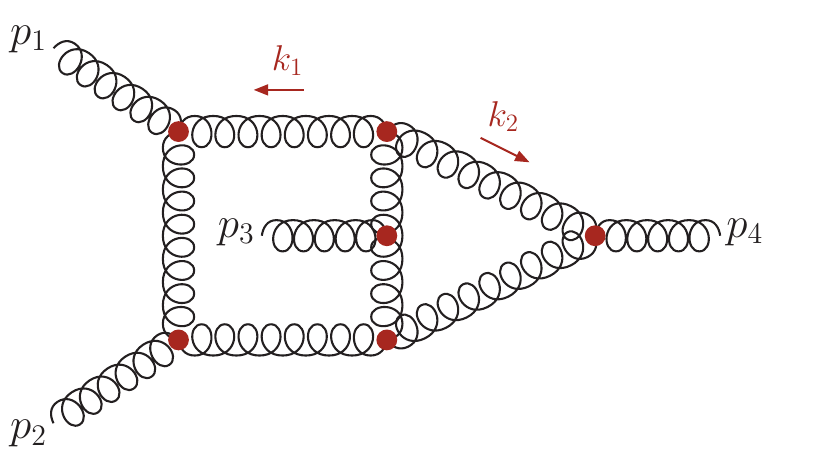}
\end{subfigure}
\caption{Feynman graphs contributing to the non-planar double-box numerator under study. There are four contributions, considering gluons and (anti-)ghosts running within the loops.}
\label{fig:nonplanar}
\end{figure}

Here we focus on the numerator of the non-planar double-box topology constructed by the seven Feynman graphs drawn in Fig.~\ref{fig:nonplanar}. The depicted graphs are part of the full-color scattering amplitude of the process $gg \to gg$. The inverse propagators describing the family of this topology are chosen as 
\begin{equation*}
    \begin{gathered}
       D_1=k_1^2, \quad D_2=\left(k_1+p_1\right){}^2, \quad D_3=\left(k_1+p_{12}\right){}^2, \quad D_4=\left(k_1+k_2-p_3\right){}^2, \quad \\ D_5=\left(k_1+k_2\right){}^2, \quad D_6=k_2^2, \quad D_7=\left(k_2-p_{123}\right){}^2, \quad D_8=\left(k_2-p_{12}\right){}^2, \quad D_9=\left(k_2-p_1\right){}^2
    \end{gathered}
\end{equation*}

We seek again to solve Eq.~(\ref{eq:4p7p}) using cut equations and the known analytic expression of the numerator. The top sector cut equations
    \begin{equation}
        D_1=D_2=D_3=D_4=D_5=D_6=D_7=0
    \end{equation}
    result in the determination of 7 invariants
    \begin{equation}
    \begin{gathered}    
        k_1\cdot k_1\to 0, \quad k_1\cdot k_2\to 0, \quad k_1\cdot p_1\to 0, \quad k_1\cdot p_2\to -\frac{s}{2},\\ k_2\cdot k_2\to 0, \quad k_2\cdot p_2\to k_1\cdot p_3-k_2\cdot p_1, \quad k_2\cdot p_3\to -\left(k_1\cdot p_3\right)
        \label{eq:7cut-np}
        \end{gathered}
    \end{equation}
    By applying the above relation on both sides of Eq.~(\ref{eq:4p7p}) we can fully determine the polynomial $P_7$.
    The polynomial $P_7$ consists of 97 coefficients over the ISP monomials $\{k_1\cdot p_3, \; k_1\cdot \eta, \; k_2\cdot p_1, \; k_2\cdot \eta\}$. 

    Subtracting the $P_7$ in Eq.~(\ref{eq:4p7p}), we can now determine the polynomials $P_6$ in the same way. There are seven six-cuts and therefore seven $P_6$ polynomials to determine. As an example, the first six-cut,
    \begin{equation}
        D_2=D_3=D_4=D_5=D_6=D_7=0
    \end{equation}
    leads to
    \begin{equation}
        \begin{gathered}
       k_1\cdot p_2\to -\frac{s}{2}, \quad k_2\cdot p_3\to -\left(k_2\cdot p_1\right)-k_2\cdot p_2, \quad k_1\cdot p_3\to k_2\cdot p_1+k_2\cdot p_2,\\ k_1\cdot k_2\to k_1\cdot p_1, \quad k_1\cdot k_1\to -2 k_1\cdot p_1, \quad k_2\cdot k_2\to 0
    \end{gathered}
\end{equation}
    whereas there are now 5 ISP $\{k_1\cdot p_1,k_1\cdot \eta ,k_2\cdot p_1,k_2\cdot p_2,k_2\cdot \eta\}$. The polynomial $P^{(6)}_1$ consists of 126 coefficients. The data for all cuts are summarized in Tab.~\ref{tab:double-box-data}. The analytic solutions for the polynomials satisfy explicitly Eq.~(\ref{eq:4p7p}).
\begin{table}[t!]
    \centering
    \begin{tabular}{|c|c|c|c|}
    \hline
        Level & Number of cuts & Number of coefficients & Scaling \\ \hline
         7 & 1 & 97 & 5,4,4 \\
         6 & 7  & 812 & 4,4,4\\
         5 & 21 & 1612 & 3,3,3\\
         4 & 35 & 922 & 2,2,2\\
         3 & 35 & 124 & 1,1,1\\
         2 & 21 & 0 & 0,0,0\\
         \hline
    \end{tabular}
    \caption{Non-planar double box linear fit information beginning with 7 cut.}
    \label{tab:double-box-data-non-planar}
\end{table}

We now seek to solve Eq.~(\ref{eq:4p9p}), namely projecting over all 9 propagators in the non-planar double-box topology. The top-level cut equations read 
\begin{equation}
        D_1=D_2=D_3=D_4=D_5=D_6=D_7=D_8=D_9=0
    \end{equation}
which leads to 
\begin{equation}
    \begin{gathered}    
        k_1\cdot k_1\to 0, \quad k_1\cdot k_2\to 0, \quad k_1\cdot p_1\to 0, \quad k_1\cdot p_2\to -\frac{s}{2}, \quad k_1\cdot p_3\to \frac{s}{2},\\ k_2\cdot k_2\to 0, \quad k_2\cdot p_1\to 0, \quad k_2\cdot p_2\to \frac{s}{2}, \quad k_2\cdot p_3\to -\frac{s}{2}
        \label{eq:9cut-np}
        \end{gathered}
    \end{equation}
    
 By applying the above relation on both sides of Eq.~(\ref{eq:4p9p}), we can fully determine the polynomial $P_9$.
    The polynomial $P_9$ consists of 15 coefficients over the ISP monomials $\{k_1\cdot \eta ,k_2\cdot \eta\}$.

     Subtracting as before the $P_9$ in Eq.~(\ref{eq:4p9p}), we can now determine the polynomials $P_8$ in the same way. There are nine 8-cuts and therefore nine $P_8$ polynomials to determine. As an example, the first 8-cut,
     \begin{equation}
        D_2=D_3=D_4=D_5=D_6=D_7=D_8=D_9=0
    \end{equation}
    leads to
    \begin{equation}
        \begin{gathered}
       k_2\cdot p_1\to 0, \quad k_2\cdot p_2\to \frac{s}{2}, \quad k_1\cdot p_2\to -\frac{s}{2}, \quad k_2\cdot p_3\to -\frac{s}{2}, \quad k_1\cdot p_3\to \frac{s}{2},\\ k_1\cdot k_2\to k_1\cdot p_1,  \quad k_1\cdot k_1\to -2 k_1\cdot p_1, \quad k_2\cdot k_2\to 0
    \end{gathered}
\end{equation}
whereas there are now 3 ISP $\{k_1\cdot p_1,k_1\cdot \eta ,k_2\cdot \eta\}$. The polynomial $P^{(8)}_1$ consists of 35 coefficients. The data for all cuts are summarized in Tab.~\ref{tab:non-planar-double-box-data-9}. The analytic solutions for the polynomials satisfy explicitly Eq.~(\ref{eq:4p9p}).
\begin{table}[t!]
    \centering
    \begin{tabular}{|c|c|c|c|}
    \hline
        Level & Number of cuts & Number of coefficients & Scaling \\ \hline
         9 & 1 & 15 & 4,4,4 \\
         8 & 9  & 267 & 4,4,4\\
         7 & 36 & 1168 & 3,3,3\\
         6 & 84 & 1601 & 2,2,2\\
         5 & 126 & 700 & 1,1,1\\
         4 & 126 & 72 & 0,0,0\\
         \hline
    \end{tabular}
    \caption{Penta-triangle linear fit information beginning with 9-cut.}
    \label{tab:non-planar-double-box-data-9}
\end{table}
The rest of the analysis, regarding the fit by cut and the global fit in $d=4-2\epsilon$ and $d=4$ dimensions, is the same as for the planar double box. 

\subsubsection{$gg\to t\bar{t}$}
\label{ttbar}

\begin{figure}[t!]
\centering
\begin{subfigure}[b]{0.328\textwidth}
\centering
\includegraphics[scale=0.36]{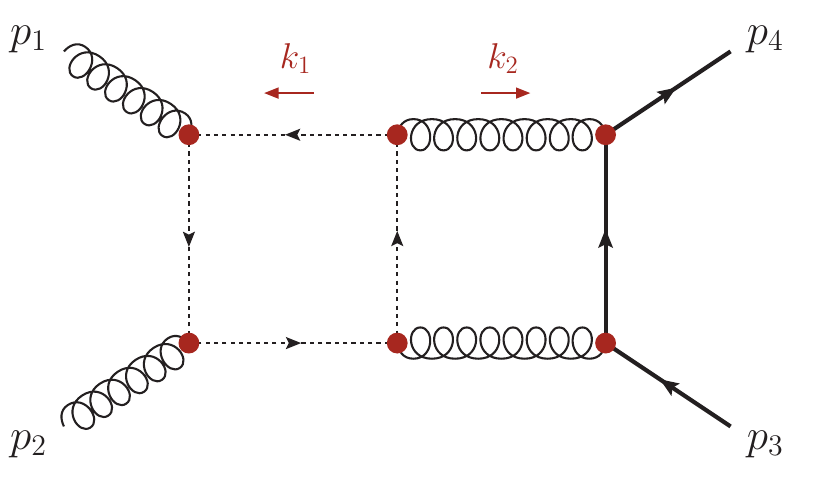}
\end{subfigure}
\hfill
\begin{subfigure}[b]{0.328\textwidth}
\centering
\includegraphics[scale=0.36]{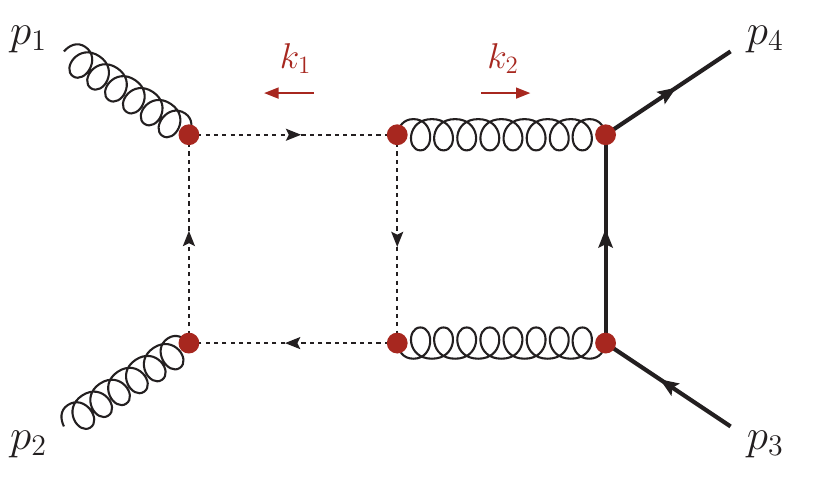}
\end{subfigure}
\hfill
\begin{subfigure}[b]{0.328\textwidth}
\centering
\includegraphics[scale=0.36]{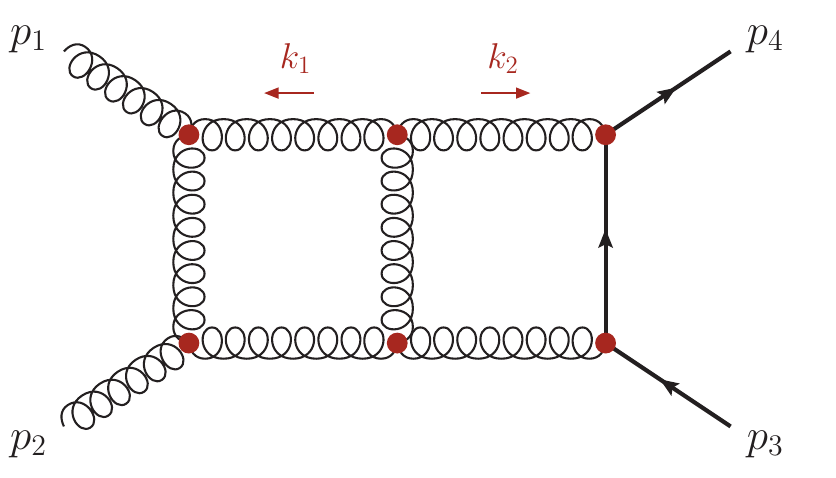}
\end{subfigure} \\
\begin{subfigure}[b]{0.328\textwidth}
\centering
\includegraphics[scale=0.46]{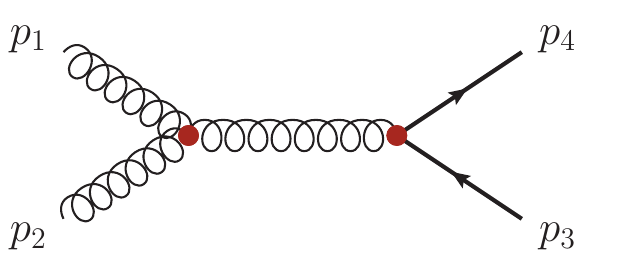}
\end{subfigure}
\caption{Feynman graphs contributing to the $gg\to t\bar{t}$ numerator under study. There are three contributions, considering gluons and (anti-)ghosts running within the $k_1$ loop, which are collected in the first line of this Figure. In the second line, we quote the tree-order graph used for the summation over polarizations. The thick black line indicates the (anti-)top quark.}
\label{fig:ttbar}
\end{figure}

In this subsection, we tackle a numerator topology consisting of one massive propagator. More specifically, we study the topology consisting of the three Feynman graphs sketched in Fig.~\ref{fig:ttbar}, which contribute to the two-loop scattering amplitude of $t\bar{t}$ production. Notice that in order to avoid unnecessary complications due to the appearance of four-dimensional spinors in the helicity amplitude at this stage of the investigation, the analytic expression of the numerator is constructed as fully summed over polarizations of the product of the two-loop contributions with the tree-order one, $\sum({\cal M}^{(0)})^*{\cal M}^{(2)}$. The propagators describing the family of this topology are chosen as 
\begin{equation*}
        \begin{gathered}
           D_1=k_1^2, \quad D_2=\left(k_1+p_1\right){}^2, \quad D_3=\left(k_1+p_{12}\right){}^2, \quad D_4=\left(k_1+k_2\right){}^2, \quad D_5=k_2^2, \\ D_6=\left(k_2-p_{123}\right){}^2-m_t^2, \quad D_7=\left(k_2-p_{12}\right){}^2, \quad D_8=\left(k_2-p_1\right){}^2, \quad D_9=\left(k_1+p_{123}\right){}^2
        \end{gathered}
\end{equation*}
The 7-cut is given by 
\begin{equation}
    \begin{gathered}
        k_1\cdot k_1\to 0, \quad k_1\cdot k_2\to 0,k_1\cdot p_1\to 0, \quad k_1\cdot p_2\to -\frac{s}{2}, \quad k_2\cdot k_2\to 0,\\ k_2\cdot p_2\to \frac{s}{2}-k_2\cdot p_1, \quad k_2\cdot p_3\to -\frac{s}{2}
    \end{gathered}
\end{equation}
and in that case the $P_7$ polynomial has no dependence on the loop momenta, $P_7=32 s^3 m_t^2 \left(2 m_t^2-s-2 t\right)$. The first 6-cut is given by 
\begin{equation}
    \begin{gathered}
       k_2\cdot p_3\to -\frac{s}{2}, \quad k_2\cdot p_1\to \frac{s}{2}-k_2\cdot p_2, \quad k_1\cdot p_2\to -\frac{s}{2},\\ k_1\cdot k_2\to k_1\cdot p_1, \quad k_2\cdot k_2\to 0, \quad k_1\cdot k_1\to -2 k_1\cdot p_1
    \end{gathered}
\end{equation}
and the corresponding $P_6$ polynomials consists of 15 monomials over $\left\{k_1\cdot p_1,k_1\cdot p_3,k_2\cdot p_2\right\}$.
The data for all cuts are summarized in Tab.~\ref{tab:ttbar-data}. The analytic solutions for the polynomials satisfy explicitly Eq.~(\ref{eq:4p7p}).
\begin{table}[t!]
    \centering
    \begin{tabular}{|c|c|c|c|}
    \hline
        Level & Number of cuts & Number of coefficients & Scaling \\ \hline
         7 & 1 & 1 & 0,0,0 \\
         6 & 7  & 80 & 3,3,3\\
         5 & 21 & 192 & 2,2,2\\
         4 & 35 & 132 & 1,1,1\\
         3 & 35 & 18 & 0,0,0\\
         2 & 21 & 0 & 0,0,0\\
         \hline
    \end{tabular}
    \caption{$gg \to t \bar{t}$ linear fit information beginning with 7-cut.}
    \label{tab:ttbar-data}
\end{table}
As is evident, for processes involving quarks, even massive, the reduction is an order of magnitude simpler than in the case of fully gluonic amplitudes. 

Projecting over all 9 propagators in this topology, the 9-cut is given by  
\begin{equation}
    \begin{gathered}
       k_1\cdot k_1\to 0, \quad k_1\cdot k_2\to 0,k_1\cdot p_1\to 0, \quad k_1\cdot p_2\to -\frac{s}{2}, \quad k_1\cdot p_3\to \frac{s}{2},\\ k_2\cdot k_2\to 0, \quad k_2\cdot p_1\to 0, \quad k_2\cdot p_2\to \frac{s}{2}, \quad k_2\cdot p_3\to -\frac{s}{2}
    \end{gathered}
\end{equation}
with $P_9$ given as before by $P_9=32 s^3 m_t^2 \left(2 m_t^2-s-2 t\right)$. The data for all cuts are summarized in Tab.~\ref{tab:ttbar-data}. The analytic solutions for the polynomials satisfy explicitly Eq.~(\ref{eq:4p9p}).
\begin{table}[t!]
    \centering
    \begin{tabular}{|c|c|c|c|}
    \hline
        Level & Number of cuts & Number of coefficients & Scaling \\ \hline
         9 & 1 & 1 & 0,0,0 \\
         6 & 9 & 17 & 2,1,2\\
         5 & 36 & 122 & 2,2,2\\
         4 & 84 & 197 & 1,1,1\\
         3 & 126 & 83 & 0,0,0\\
         4 & 126 & 0 & 0,0,0\\
         \hline
    \end{tabular}
    \caption{$gg \to t \bar{t}$ linear fit information beginning with 9-cut.}
    \label{tab:ttbar-data9}
\end{table}
In both cases, the rest of the analysis works along the same lines as described for the double-box presented above.

\subsection{Five-Point Kinematics}
\label{Five_point}
\subsubsection{Penta-box topologies}
\label{Pentabox}

\begin{figure}[t!]
\centering
\begin{subfigure}[b]{0.328\textwidth}
\centering
\includegraphics[scale=0.38]{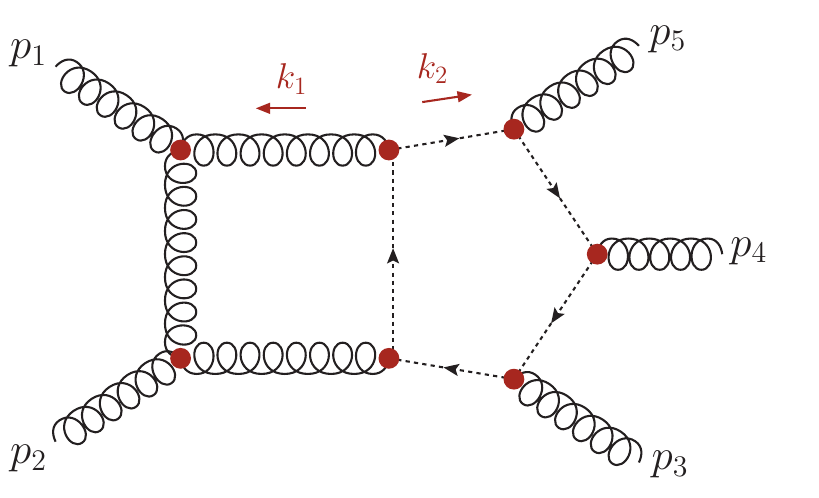}
\end{subfigure}
\hfill
\begin{subfigure}[b]{0.328\textwidth}
\centering
\includegraphics[scale=0.38]{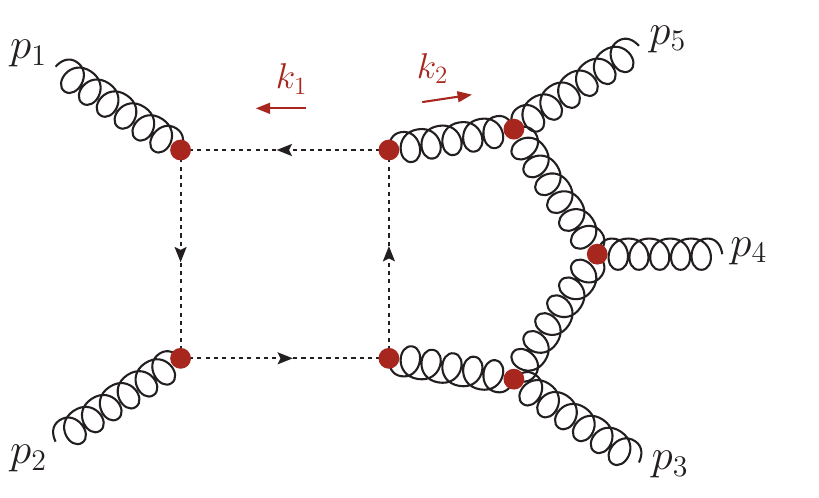}
\end{subfigure}
\hfill
\begin{subfigure}[b]{0.328\textwidth}
\centering
\includegraphics[scale=0.38]{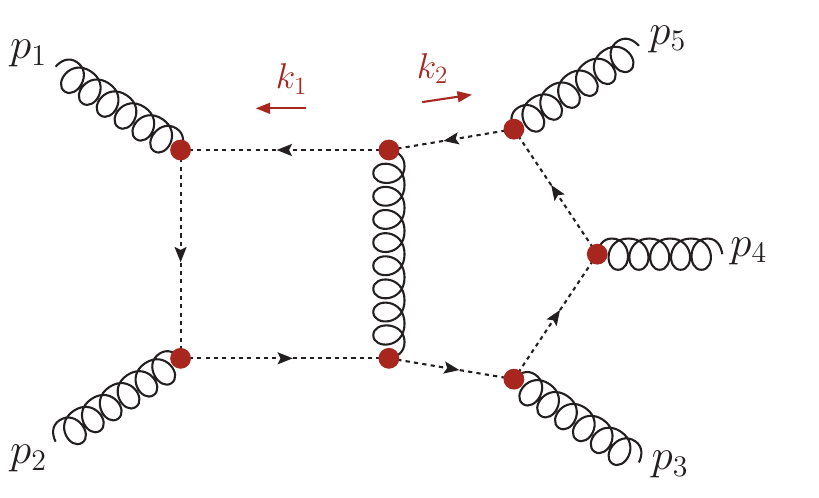}
\end{subfigure}
\begin{subfigure}[b]{0.328\textwidth}
\centering
\includegraphics[scale=0.38]{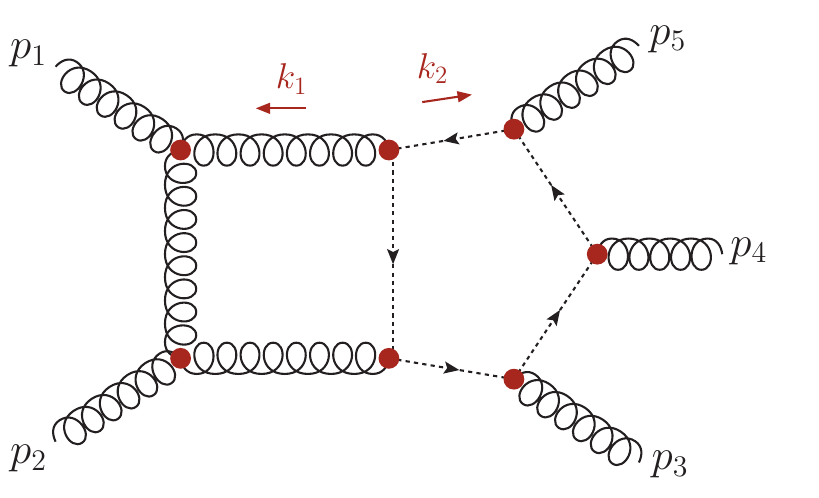}
\end{subfigure}
\hfill
\begin{subfigure}[b]{0.328\textwidth}
\centering
\includegraphics[scale=0.38]{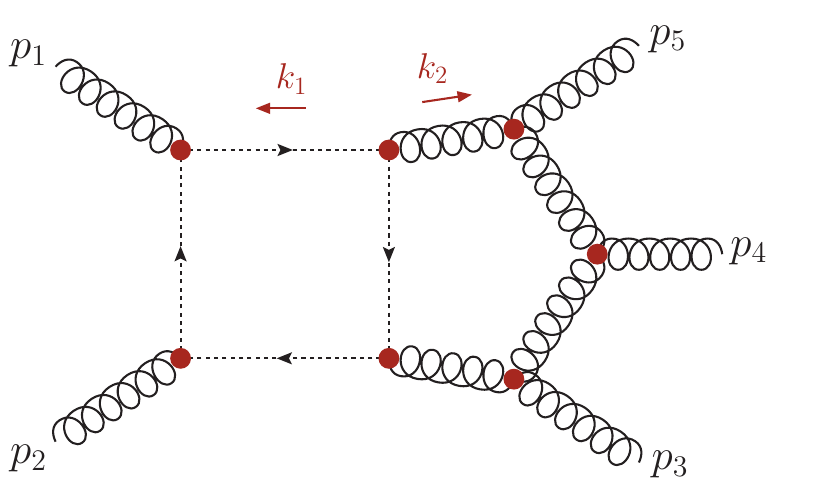}
\end{subfigure}
\hfill
\begin{subfigure}[b]{0.328\textwidth}
\centering
\includegraphics[scale=0.38]{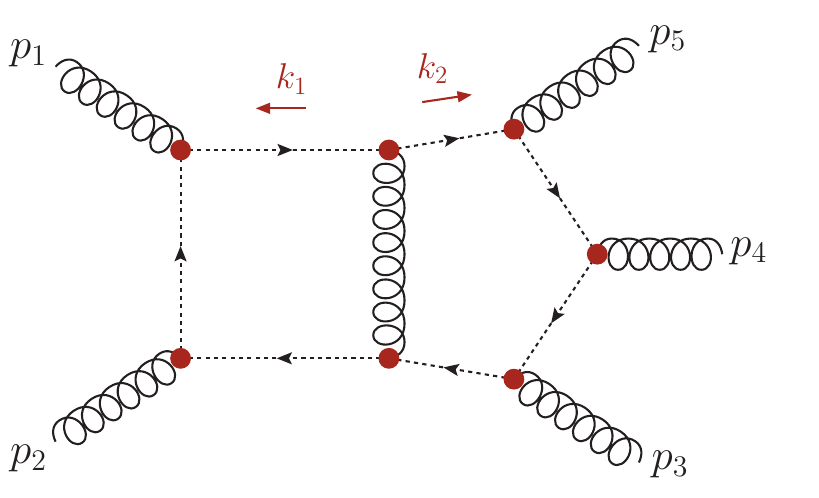}
\end{subfigure}
\hfill
\begin{subfigure}[b]{0.328\textwidth}
\centering
\includegraphics[scale=0.38]{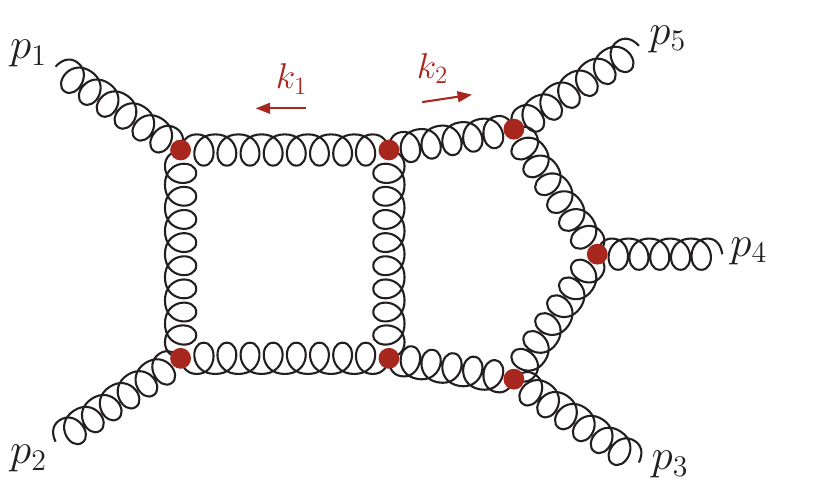}
\end{subfigure}
\caption{Feynman graphs contributing to the penta-box numerator under study. There are seven contributions, considering gluons and (anti-)ghosts running within the loops.}
\label{fig:pentabox}
\end{figure}

Here we consider the numerator of the penta-box topology constructed by the seven Feynman graphs depicted in Fig.~\ref{fig:pentabox}, which contributes to the scattering amplitude of the process $gg \to ggg$. For the analysis presented below, we have used the following helicity assignment: $\lambda_1=+1,\lambda_2=+1,\lambda_3=-1,\lambda_4=-1,\lambda_5=-1$, for the helicities of the incoming gluons. Our choice for the inverse propagators describing the family of this topology is the following
    \begin{equation}
        \begin{gathered}
            D_1=k_1^2, \quad D_2=\left(k_1+p_1\right){}^2, \quad D_3=\left(k_1+p_{12}\right){}^2, \quad D_4=\left(k_1+k_2\right){}^2, \\  D_5=k_2^2, \quad D_6=\left(k_2-p_{1234}\right){}^2, \quad D_7=\left(k_2-p_{123}\right){}^2, \quad D_8=\left(k_2-p_{12}\right){}^2, \\ D_9=\left(k_1+p_{123}\right){}^2, \quad D_{10}=\left(k_1+p_{1234}\right){}^2, \quad D_{11}=\left(k_2-p_1\right){}^2
        \end{gathered}
    \label{eq:5p-planar-family}
    \end{equation}
     The external kinematics is described by 4 independent momenta and five independent invariants, which can be chosen to be $S_5=\{s_{12},s_{23},s_{34},s_{45},s_{15}\}$, where $s_{ij}=(p_i+p_j)^2$. In order to proceed with the reduction in an analytic setup, the expression of the numerator needs to be expressed in terms of the 11 invariants $V_{11}$, defined as the set of variables $k_i\cdot k_j,$ with $i,j=1,2$, and $k_i\cdot p_j$ with $i=1,2,\; j=1,\ldots,4$. Since the analytic expression of the numerator involves the polarization vectors of the external gluons, $\varepsilon_i,i=1,\ldots,5$, scalar products of the form $k_i\cdot\varepsilon_j$ and $p_i\cdot\varepsilon_j$ need to be expressed in terms of the set of variables $S_5$ and $V_{11}$. To this end, the following relation is used
     \begin{equation}
         q_i\cdot q_j=G^{-1}_{kl} \, q_i\cdot p_k \, q_j \cdot p_l 
     \label{eq:gram}
     \end{equation}
     where $q_i$ stands for any momentum or polarization vector and $G$ denotes the Gram matrix, $G_{ij}=p_i\cdot p_j,i,j=1,\ldots,4$ expressed in terms of the $S_5$ variables.
     Barring that the analytic expression for the inverse Gram matrix and the numerator are complicated expressions, it is convenient to work in a numerical setup, using exact arithmetic. The numerical values for the variables $S_5$, are chosen as $\left\{s_{12}\to 1,s_{34}\to \frac{1}{4},s_{45}\to \frac{1}{4},s_{15}\to -\frac{1}{4},s_{23}\to -\frac{1}{8}\right\}$. The polarization vectors are defined by
     \begin{equation}
     \begin{gathered}
         \varepsilon_\mu^{+}(p_i)=\frac{1}{\sqrt{2}\bar{u}_-(p_{i+1})u_+(p_i)}\bar{u}_-(p_{i+1})\gamma_\mu u_-(p_i)\\
          \varepsilon_\mu^{-}(p_i)=-\frac{1}{\sqrt{2}\bar{u}_+(p_{i+1})u_-(p_i)}\bar{u}_+(p_{i+1})\gamma_\mu u_+(p_i)
          \end{gathered}
          \label{eq:pol-vextors}
     \end{equation}
     for $i=1,\ldots,5$, where in the above formula the following the $p_6\to p_1$ identification is assumed. With the above identifications, the numerator consists of monomials composed of the $V_{11}$ variables, with exact numerical coefficients. 

     We seek to solve Eq.~(\ref{eq:5p8p}). The maximal cut is given by 
     \begin{equation}
         \begin{gathered}
             k_1\cdot k_1\to 0, \quad k_1\cdot k_2\to 0, \quad k_1\cdot p_1\to 0, \quad k_1\cdot p_2\to -\frac{s_{12}}{2}, \quad k_2\cdot k_2\to 0,\\ k_2\cdot p_2\to \frac{s_{12}}{2}-k_2\cdot p_1, \quad k_2\cdot p_3\to \frac{s_{45}}{2}-\frac{s_{12}}{2}, \quad k_2\cdot p_4\to -\frac{s_{45}}{2}
         \end{gathered}
     \end{equation}
     The $P_8$ polynomial consists of 50 terms, composed of monomials in the ISP variables $\left\{k_1\cdot p_3,k_1\cdot p_4,k_2\cdot p_1\right\}$. By following the usual subtraction procedure, the solution for all polynomials is found to satisfy Eq.~(\ref{eq:5p8p}). The data of this solution are given in Tab.~\ref{tab:pentabox-data}.
     \begin{table}[t!]
    \centering
    \begin{tabular}{|c|c|c|c|}
    \hline
        Level & Number of cuts & Number of coefficients & Scaling \\ \hline
         8 & 1 & 50 & 4,5,5 \\
         7 & 8  & 705 & 4,4,5\\
         6 & 28 & 2550 & 4,4,4\\
         5 & 56 & 3508 & 3,3,3\\
         4 & 70 & 1902 & 2,2,2\\
         3 & 56 & 348 & 1,1,1\\
         2 & 28 & 12 & 0,0,0\\
         \hline
    \end{tabular}
    \caption{Penta-box linear fit information beginning with 8-cut.}
    \label{tab:pentabox-data}
\end{table}
     Seeking now to solve Eq.~(\ref{eq:5p11p}) the maximal cut is given by 
     \begin{equation}
         \begin{gathered}
             k_1\cdot k_1\to 0, \quad  k_1\cdot k_2\to 0, \quad k_1\cdot p_1\to 0, \quad k_1\cdot p_2\to -\frac{s_{12}}{2}, \quad k_1\cdot p_3\to \frac{s_{12}}{2}-\frac{s_{45}}{2},\\ k_1\cdot p_4\to \frac{s_{45}}{2}, \quad k_2\cdot k_2\to 0, \quad k_2\cdot p_1\to 0, \quad k_2\cdot p_2\to \frac{s_{12}}{2}, \\ k_2\cdot p_3\to \frac{s_{45}}{2}-\frac{s_{12}}{2}, \quad k_2\cdot p_4\to -\frac{s_{45}}{2}
         \end{gathered}
     \end{equation}
     and as before, the cut data are shown in Tab.~\ref{tab:pentabox-data11}.
     \begin{table}[t!]
    \centering
    \begin{tabular}{|c|c|c|c|}
    \hline
        Level & Number of cuts & Number of coefficients & Scaling \\ \hline
         11 & 1 & 1 & 0,0,0 \\
         10 & 11  & 47 & 3,4,4\\
         9 & 55 & 502 & 3,4,4\\
         8 & 165 & 2313 & 3,3,3\\
         7 & 330 & 3715 & 2,2,2\\
         6 & 462 & 2255 & 1,1,1\\
         5 & 462 & 425 & 0,0,0\\
         \hline
    \end{tabular}
    \caption{Penta-box linear fit information beginning with 11-cut.}
    \label{tab:pentabox-data11}
\end{table}
The fit by cut and the global fit in $d=4-2\epsilon$ dimensions is realized as before through the numerical representation of the loop momenta given in section~\ref{fit by cut}.
    
\subsubsection{$t\bar{t}H$}
\label{ttbarH}

\begin{figure}[t!]
\centering
\begin{subfigure}[b]{0.328\textwidth}
\centering
\includegraphics[scale=0.38]{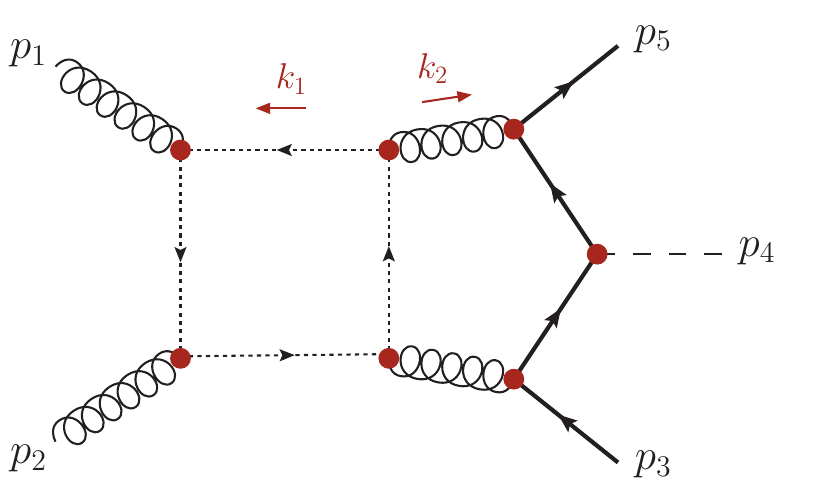}
\end{subfigure}
\hfill
\begin{subfigure}[b]{0.328\textwidth}
\centering
\includegraphics[scale=0.38]{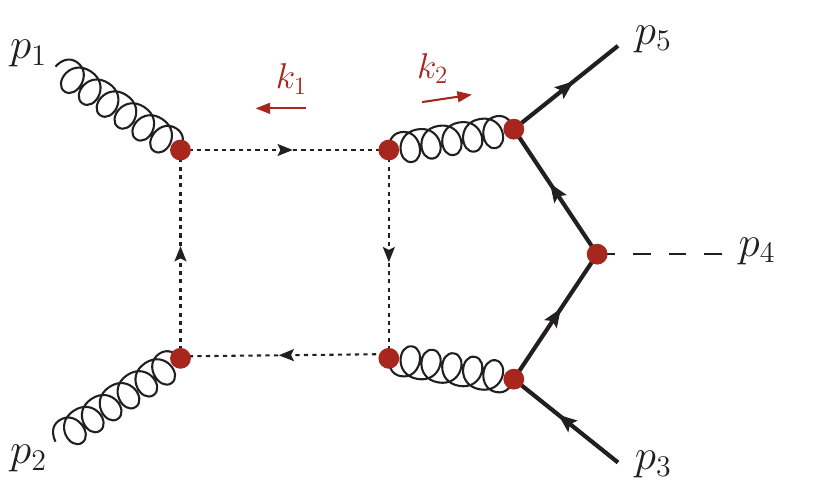}
\end{subfigure}
\hfill
\begin{subfigure}[b]{0.328\textwidth}
\centering
\includegraphics[scale=0.38]{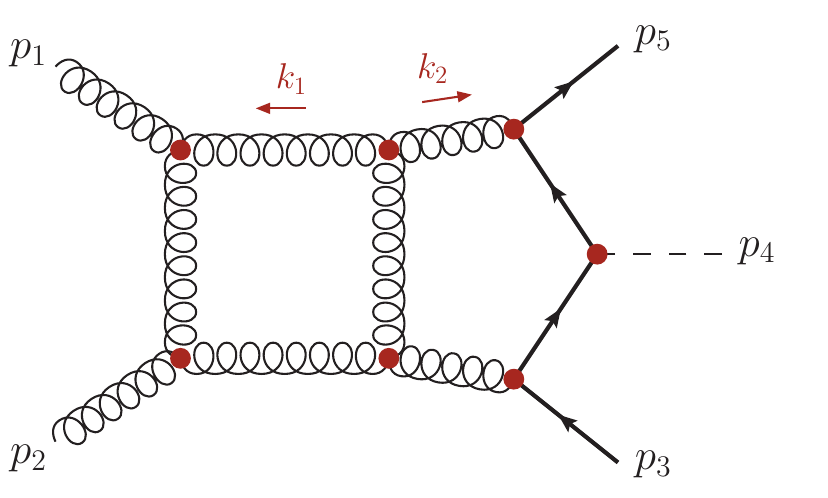}
\end{subfigure} \\
\begin{subfigure}[b]{0.328\textwidth}
\centering
\includegraphics[scale=0.38]{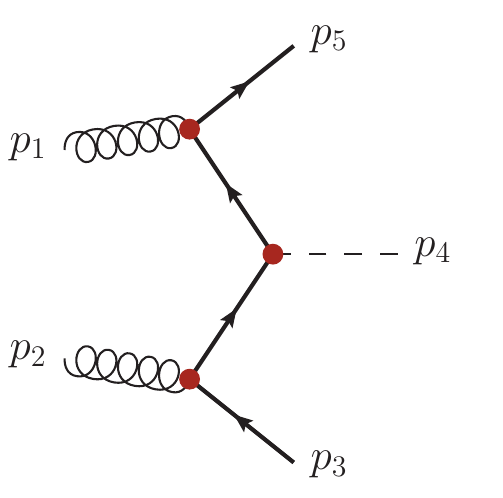}
\end{subfigure}
\caption{Feynman graphs contributing to the $gg\to t\bar{t} H$ numerator under study. There are three contributions, considering gluons and (anti-)ghosts running within the $k_1$ loop, which are collected in the first line of this Figure. In the second line, we quote the tree-order graph used for the summation over polarizations. The dashed line indicates the Higgs boson.}
\label{fig:ttH}
\end{figure}

Herein, we study a numerator topology consisting of two massive propagators.  This corresponds to the three graphs depicted in Fig.~\ref{fig:ttH}, which contribute to the scattering amplitude of the process $gg\to ttH$. We again note that in order to avoid unnecessary complications due to the appearance in the helicity amplitude of four-dimensional spinors, the analytic expression of the numerator is constructed as a fully summed over polarizations  of the product of the two-loop contributions with the tree-order one.
The inverse propagators describing the family for this topology are chosen as 
    \begin{equation}
        \begin{gathered}
           D_1=k_1^2, \quad D_2=\left(k_1+p_1\right){}^2, \quad D_3=\left(k_1+p_{12}\right){}^2, \quad D_4=\left(k_1+k_2\right){}^2, \quad D_5=k_2^2, \\ D_6=\left(k_2-p_{1234}\right){}^2-m_t^2, \quad D_7=\left(k_2-p_{123}\right){}^2-m_t^2, \quad D_8=\left(k_2-p_{12}\right){}^2, \\ D_9=\left(k_1+p_{123}\right){}^2, \quad D_{10}=\left(k_1+p_{1234}\right){}^2, \quad D_{11}=\left(k_2-p_1\right){}^2      \end{gathered}
    \end{equation}
    \label{eq:ttHr-planar-family}
    The 8-cut is given by 
\begin{equation}
    \begin{gathered}
       k_1\cdot k_1\to 0, \quad k_1\cdot k_2\to 0, \quad k_1\cdot p_1\to 0, \quad k_1\cdot p_2\to -\frac{s_{12}}{2}, \quad k_2\cdot k_2\to 0,\\ k_2\cdot p_2\to \frac{s_{12}}{2}-k_2\cdot p_1, \quad k_2\cdot p_3\to \frac{1}{2} \left(-m_t^2-s_{12}+s_{45}\right), \quad k_2\cdot p_4\to \frac{1}{2} \left(m_t^2-s_{45}\right)
    \end{gathered}
\end{equation}
and in that case, the $P_8$ polynomial consists of 32 terms composed of monomials in the ISP variables $k_1\cdot p_3,k_1\cdot p_4,k_2\cdot p_1$. 
The data for all cuts are summarized in Tab.~\ref{tab:ttH-data}. The analytic solutions for the polynomials satisfy explicitly Eq.~(\ref{eq:4p7p}).
\begin{table}[t!]
    \centering
    \begin{tabular}{|c|c|c|c|}
    \hline
        Level & Number of cuts & Number of coefficients & Scaling \\ \hline
         8 & 1 & 32 & 4,3,5 \\
         7 & 8  & 293 & 4,3,4\\
         6 & 28 & 735 & 3,2,2\\
         5 & 56 & 651 & 2,2,2\\
         4 & 70 & 181 & 1,1,1\\
         3 & 56 & 8 & 0,0,0\\
         \hline
    \end{tabular}
    \caption{$gg \to t\bar{t} H$ linear fit information beginning  with 8-cut.}
    \label{tab:ttH-data}
\end{table}
Projecting over all 11 propagators in this topology, the 11-cut is given by  
\begin{equation}
    \begin{gathered}
       k_1\cdot k_1\to 0, \quad k_1\cdot k_2\to 0, \quad k_1\cdot p_1\to 0, \quad k_1\cdot p_2\to -\frac{s_{12}}{2}, \quad k_1\cdot p_3\to \frac{1}{2} \left(s_{12}-s_{45}\right),\\ k_1\cdot p_4\to \frac{1}{2} \left(s_{45}-m_t^2\right), \quad k_2\cdot k_2\to 0, \quad k_2\cdot p_1\to 0, \quad k_2\cdot p_2\to \frac{s_{12}}{2},\\ k_2\cdot p_3\to \frac{1}{2} \left(-m_t^2-s_{12}+s_{45}\right), \quad k_2\cdot p_4\to \frac{1}{2} \left(m_t^2-s_{45}\right)
    \end{gathered}
\end{equation}
with $P_{11}$ consisting of a single term expressed as a function of the external invariants $\left\{m_H,m_t,s_{12},s_{15},s_{23},s_{34},s_{45}\right\}$. The data for all cuts are summarized in Tab.~\ref{tab:ttH-data11}. The analytic solutions for the polynomials satisfy explicitly Eq.~(\ref{eq:4p9p}).
\begin{table}[t!]
    \centering
    \begin{tabular}{|c|c|c|c|}
    \hline
        Level & Number of cuts & Number of coefficients & Scaling \\ \hline
         11 & 1 & 1 & 0,0,0 \\
         10 & 11 & 32 & 2,2,2\\
         9 & 55 & 270 & 2,2,3\\
         8 & 165 & 766 & 2,2,2\\
         7 & 330 & 734 & 1,1,1\\
         6 & 462 & 158 & 0,0,0\\
         \hline
    \end{tabular}
    \caption{$gg \to t\bar{t} H$ linear fit information beginning  with 11-cut.}
    \label{tab:ttH-data11}
\end{table}
In both cases, the rest of the analysis works in the same lines as for the double-box presented above.

\subsection{Six-Point Kinematics}
\label{Six_point}
\subsubsection{Six-gluon topology}
\label{6-gluon}

\begin{figure}[t!]
\centering
\includegraphics[scale=0.45]{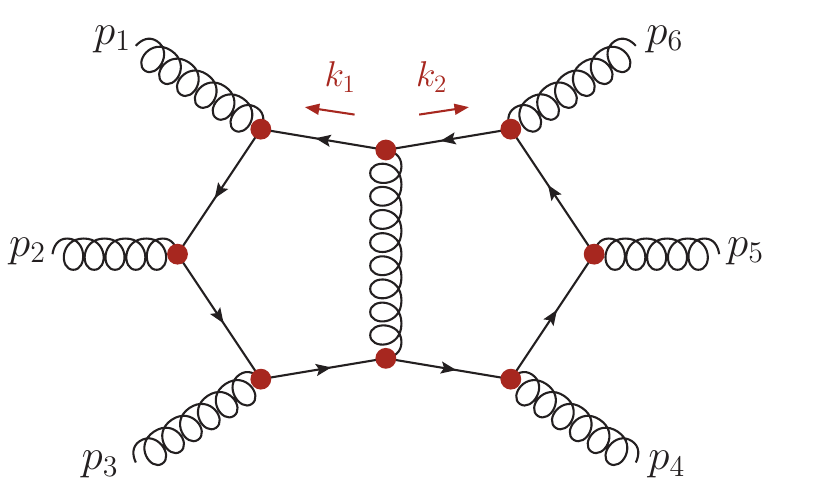}
\caption{Feynman graph contributing to the six-gluon numerator under study. The normal line is used for denoting a massless quark.}
\label{fig:sixphoton}
\end{figure}

In this subsection, we apply our method to the case of the six-gluon two-loop numerator topology of Fig.~\ref{fig:sixphoton}. This graph is part of the $gg\to gggg$ scattering amplitude. For the analysis presented below, we have used the following helicity assignment: $\lambda_1=+1,\lambda_2=+1,\lambda_3=-1,\lambda_4=-1,\lambda_5=-1,\lambda_6=-1$, for the helicities of the incoming gluons. We define the propagators of the family in which this topology belongs, as
 \begin{equation*}
        \begin{gathered}
           D_1=k_1^2, \quad D_2=\left(k_1+p_1\right){}^2, \quad D_3=\left(k_1+p_{12}\right){}^2, \quad D_4=\left(k_1+p_{123}\right){}^2, \quad D_5=\left(k_1+k_2\right){}^2,\\  D_6=k_2^2, \quad D_7=\left(k_2-p_{12345}\right){}^2, \quad  D_8=\left(k_2-p_{1234}\right){}^2, \quad D_9=\left(k_2-p_{123}\right){}^2, \\ D_{10}=\left(k_1+p_{1234}\right){}^2, \quad D_{11}=\left(k_1+p_{12345}\right){}^2, \quad D_{12}=\left(k_2-p_{12}\right){}^2, \quad D_{13}=\left(k_2-p_1\right){}^2
        \end{gathered}
    \label{eq:6p-planar-family}
\end{equation*}
Since out of the six external momenta only four are independent in $d=4$ dimensions, not all 13 propagators in Eq.~(\ref{eq:6p-planar-family}) are independent. In general, for any $n$-point amplitude, with $n\ge 5$, only 11 propagators are independent. We have chosen the following subset, $\left\{D_1,\ldots,D_{10},D_{13}\right\}$ having expressed $p_5$ in terms of $p_1,\ldots,p_4$ through Eq.~(\ref{eq:gram}). As in the case of the penta-box, section~\ref{Pentabox}, the analytic expressions are hardly manageable, and it is convenient to work in a numerical setup, using exact arithmetic. The numerical values of the invariants are chosen as~\footnote{Notice that for 4-dimensional external momenta only 8 invariants are independent. In fact, $s_{345}$, for instance, can be determined by the requirement that the five-momenta Gram determinant, $G(12345)$, vanishes, see also ref.~\cite{Abreu:2024fei}.}
\begin{equation*}
\left\{s_{12}\to 4,s_{23}\to -1,s_{34}\to 1,s_{45}\to \frac{5}{4},s_{56}\to \frac{1}{2},s_{16}\to -1,s_{123}\to 2,s_{234}\to -1,s_{345}\to \frac{31}{12}\right\},
\end{equation*} 
with $s_{ijk}=(p_i+p_j+p_k)^2$. For the polarization vectors we follow Eq.~(\ref{eq:pol-vextors}), with the identification of $p_7\to p_1$. The 9-cut 
\begin{equation}
    D_1=\ldots=D_9=0
\end{equation}
is given, in the numerical point chosen, by
\begin{equation}
\begin{gathered}
    k_1\cdot k_1\to 0, \quad k_1\cdot k_2\to 0, \quad k_1\cdot p_1\to 0, \quad k_1\cdot p_2\to -2, \quad k_1\cdot p_3\to 1, \quad k_2\cdot k_2\to 0,\\ k_2\cdot p_2\to \frac{5}{2}-\frac{5 k_2\cdot p_1}{3}, \quad k_2\cdot p_3\to \frac{2 k_2\cdot p_1}{3}-\frac{3}{2}, \quad k_2\cdot p_4\to -\frac{3}{4}
\end{gathered}
\end{equation}
The reduction data are given in Tab.~\ref{tab:6p-data}.
\begin{table}[t!]
    \centering
    \begin{tabular}{|c|c|c|c|}
    \hline
        Level & Number of cuts & Number of coefficients & Scaling \\ \hline
         9 & 1 & 21 & 4,4,6 \\
         8 & 9 & 355 & 4,4,6\\
         7 & 36 & 1949 & 4,4,5\\
         6 & 84 & 4462 & 4,4,4\\
         5 & 126 & 4540 & 3,3,3\\
         4 & 126 & 2016 & 2,2,2\\
         3 & 84 & 334 & 1,1,1\\
         2 & 36 & 16 & 0,0,0\\
         \hline
    \end{tabular}
    \caption{6 gluon linear fit information beginning with 9-cut.}
    \label{tab:6p-data}
\end{table}
Projecting over the set of 11 propagators, referred above, the 11-cut is given by
\begin{equation}
    \begin{gathered}
        k_1\cdot k_1\to 0, \quad k_1\cdot k_2\to 0, \quad k_1\cdot p_1\to 0, \quad k_1\cdot p_2\to -2, \quad k_1\cdot p_3\to 1,\\ k_1\cdot p_4\to \frac{3}{4}, \quad k_2\cdot k_2\to 0, \quad k_2\cdot p_1\to 0, \quad k_2\cdot p_2\to \frac{5}{2}, \\ k_2\cdot p_3\to -\frac{3}{2}, \quad k_2\cdot p_4\to -\frac{3}{4}
    \end{gathered}
\end{equation}
and the reduction data are summarized in Tab.~\ref{tab:6p-data11}.
\begin{table}[ht]
    \centering
    \begin{tabular}{|c|c|c|c|}
    \hline
        Level & Number of cuts & Number of coefficients & Scaling \\ \hline
         11 & 1 & 1 & 0,0,0 \\
         10 & 11 & 41 & 3,3,3\\
         9 & 55 & 505 & 3,3,5\\
         8 & 165 & 2365 & 3,3,4\\
         7 & 330 & 4780 & 3,3,3\\
         6 & 462 & 4290 & 2,2,2\\
         5 & 462 & 1592 & 1,1,1\\
         4 & 330 & 200 & 0,0,0\\
         \hline
    \end{tabular}
    \caption{6 gluon linear fit information beginning with 11-cut.}
    \label{tab:6p-data11}
\end{table}

\section{Summary and Discussion}
\label{Conclusions}

In this paper, we have studied the integrand-level reduction at two loops, solving Eq.~(\ref{integrand reduction}) and Eq.~(\ref{integrand reduction over family}). Let us briefly summarize our findings:
\begin{itemize}
    \item We solved the reduction equation by two generic methods, namely, fit by cut and global fit. In the fit by cut approach, we constructed the solutions to the cut equations Eq.~(\ref{cuteqn}) and then used them to fit the polynomials $P$ appearing in Eq.~(\ref{integrand reduction}).
    \item A novel element of our studies is the observation that the reduction equation can be extended to include all inverse propagators characterizing a given family and not just the ones appearing in the Feynman graphs under consideration, see Eq.~(\ref{integrand reduction over family}). This way, the polynomials depend on a smaller number of variables. Moreover, this approach allows us to use the solutions of the cut equations for all Feynman graphs under the same family.
    \item We applied our method in both $d=4-2\epsilon$ and $d=4$ dimensions. In the $d=4-2\epsilon$ case, the numerator is expressed as a function of the four-dimensional external momenta and polarization vectors and the four-dimensional part of the loop momenta along with the $\mu_{ij}$ terms. In all cases, this is readily possible, since the numerator has been known analytically. Cut equations admit a unique solution in terms of the ISP and can be used to fit the coefficients of the polynomials either analytically or numerically.
    \item In $d=4$ dimensions, the cut equations do not admit, in general, a unique solution in terms of ISP. 
    We showed, nevertheless, how to construct solutions to both Eq.~(\ref{integrand reduction}) and Eq.~(\ref{integrand reduction over family}).
    \item In the global fit approach, both in $d=4-2\epsilon$ and $d=4$ dimensions, solutions to the cut equations are not necessary, but determining the coefficients of the polynomials involves solving linear equations with relatively large matrices. It is though a plausible alternative to the fit by cut approach, especially in validating Eq.~(\ref{integrand reduction}) and Eq.~(\ref{integrand reduction over family}).  
   \item We have successfully studied the reduction equations for several cases in 4$-$, 5$-$ and 6$-$particle topologies, with massless and massive external particles and internal propagators: double-box, penta-triangle, hexa-bubble, non-planar double-box, $t\bar t$, penta-box, $t\bar t H$ and 6-gluon with a closed quark loop. We conclude that two-loop reduction at the integrand level is within reach for arbitrary processes. We plan to implement the solutions of the cut equations in a numerical setup~\cite{Ossola:2007ax}, in order to allow for a fully-fledged numerical reduction of two-loop integrand numerators.      
\end{itemize}
A crucial element in the direction of the construction of a generic automated two-loop amplitude computational framework based on the integrand-level reduction is the calculation of the dimensionally regulated amplitudes. As we have seen in this paper, the numerical evaluation of the integrands needs to address the dependence on $\mu_{ij}$ and $\epsilon$ terms. This issue is currently under investigation~\cite{Bevilacqua:2024fec}, and we plan to publish our results in the near future~\cite{new1}. 

In order to provide useful results for phenomenological applications, the two-loop Feynman integrals in Eq.~(\ref{integral reduction}) need to be provided. At this time, a fully numerical computation of all two-loop integrals~\cite{Heinrich:2023til, Liu:2022chg} seems quite challenging. Analytic results, on the other hand, are available for master integrals in more and more cases over the recent years~\cite{Papadopoulos:2015jft, Gehrmann:2018yef, Chicherin:2018mue, Chicherin:2018old, Abreu:2020jxa, Canko:2020ylt, Chicherin:2020oor, Abreu:2021smk, Chicherin:2021dyp, Kardos:2022tpo, Badger:2022hno, FebresCordero:2023pww, Abreu:2023rco, Abreu:2024yit, Badger:2024fgb, Abreu:2024fei, Henn:2025xrc, Becchetti:2025qlu, Bargiela:2025nqc}, but IBP solutions in expressing Feynman integrals in terms of master integrals are necessary as well. In principle, this can be achieved in a process-independent way by providing appropriate IBP tables that express the Feynman integrals of Eq.~(\ref{integral}) in terms of currently known MIs. We plan to provide a proof of concept of a complete computational framework in the near future~\cite{new2}, combining {\tt HELAC-2LOOP} amplitude generator for the numerical evaluation of the numerators in $d=4-2\epsilon$ dimensions, the numerical implementation of the reduction equation, Eq.~(\ref{integrand reduction over family}), and the interface to evaluating the currently known Feynman integrals appearing in Eq.(~\ref{integral reduction}), including a detailed analysis of numerical accuracy and performance, paving the way to automated two-loop calculations.

\acknowledgments

We thank Simon Badger, Hjalte Frellesvig, Adam Kardos, Tiziano Peraro, Vasily Sotnikov, and Yang Zhang for useful discussions. The work of G.~Bevilacqua, C.~G.~Papadopoulos, and A.~Spourdalakis is supported by the Hellenic Foundation for Research and Innovation (H.F.R.I.) under the "2nd Call for H.F.R.I. Research Projects to support Faculty Members \& Researchers" (Project Number: 02674 HOCTools-II). The work of D.~Canko is supported by the European Research Council (ERC) under the European Union’s Horizon Europe research and innovation program grant agreement 101040760, \textit{High-precision multi-leg Higgs and top physics with finite fields} (ERC Starting Grant FFHiggsTop). C.~G.~Papadopoulos would like to thank the Department of Theoretical Physics at CERN and the Aspen Center for Physics and the Simons Foundation, for their kind hospitality.


\bibliographystyle{JHEP}
\bibliography{biblio}

\providecommand{\href}[2]{#2}\begingroup\raggedright\begin{thebibliography}{100}

\bibitem{Dainese:2019rgk}
A.~Dainese, M.~Mangano, A.~B. Meyer, A.~Nisati, G.~Salam, and M.~A. Vesterinen,
  eds., {\em {Report on the Physics at the HL-LHC,and Perspectives for the
  HE-LHC}}, vol.~7/2019 of {\em CERN Yellow Reports: Monographs}.
\newblock CERN, Geneva, Switzerland, 2019.

\bibitem{Caola:2022ayt}
F.~Caola, W.~Chen, C.~Duhr, X.~Liu, B.~Mistlberger, F.~Petriello, G.~Vita, and
  S.~Weinzierl, {\it {The Path forward to N$^3$LO}},  in {\em {Snowmass 2021}},
  3, 2022.
\newblock \href{http://arxiv.org/abs/2203.06730}{{\tt arXiv:2203.06730}}.

\bibitem{CMS:2025hfp}
{\bf CMS} Collaboration, {\it {Highlights of the HL-LHC physics projections by
  ATLAS and CMS}},  \href{http://arxiv.org/abs/2504.00672}{{\tt
  arXiv:2504.00672}}.

\bibitem{FCC:2025lpp}
{\bf FCC} Collaboration, M.~Benedikt et~al., {\it {Future Circular Collider
  Feasibility Study Report: Volume 1, Physics, Experiments, Detectors}},
  \href{http://arxiv.org/abs/2505.00272}{{\tt arXiv:2505.00272}}.

\bibitem{Huss:2025nlt}
A.~Huss, J.~Huston, S.~Jones, M.~Pellen, and R.~R\"ontsch, {\it {Les Houches
  2023 -- Physics at TeV Colliders: Report on the Standard Model Precision
  Wishlist}},  \href{http://arxiv.org/abs/2504.06689}{{\tt arXiv:2504.06689}}.

\bibitem{Badger:2017jhb}
S.~Badger, C.~Br\o{}nnum-Hansen, H.~B. Hartanto, and T.~Peraro, {\it {First
  look at two-loop five-gluon scattering in QCD}},  {\em Phys. Rev. Lett.} {\bf
  120} (2018), no.~9 092001, [\href{http://arxiv.org/abs/1712.02229}{{\tt
  arXiv:1712.02229}}].

\bibitem{Badger:2018enw}
S.~Badger, C.~Br\o{}nnum-Hansen, H.~B. Hartanto, and T.~Peraro, {\it {Analytic
  helicity amplitudes for two-loop five-gluon scattering: the single-minus
  case}},  {\em JHEP} {\bf 01} (2019) 186,
  [\href{http://arxiv.org/abs/1811.11699}{{\tt arXiv:1811.11699}}].

\bibitem{Abreu:2018jgq}
S.~Abreu, F.~Febres~Cordero, H.~Ita, B.~Page, and V.~Sotnikov, {\it {Planar
  Two-Loop Five-Parton Amplitudes from Numerical Unitarity}},  {\em JHEP} {\bf
  11} (2018) 116, [\href{http://arxiv.org/abs/1809.09067}{{\tt
  arXiv:1809.09067}}].

\bibitem{Abreu:2018zmy}
S.~Abreu, J.~Dormans, F.~Febres~Cordero, H.~Ita, and B.~Page, {\it {Analytic
  Form of Planar Two-Loop Five-Gluon Scattering Amplitudes in QCD}},  {\em
  Phys. Rev. Lett.} {\bf 122} (2019), no.~8 082002,
  [\href{http://arxiv.org/abs/1812.04586}{{\tt arXiv:1812.04586}}].

\bibitem{Abreu:2019odu}
S.~Abreu, J.~Dormans, F.~Febres~Cordero, H.~Ita, B.~Page, and V.~Sotnikov, {\it
  {Analytic Form of the Planar Two-Loop Five-Parton Scattering Amplitudes in
  QCD}},  {\em JHEP} {\bf 05} (2019) 084,
  [\href{http://arxiv.org/abs/1904.00945}{{\tt arXiv:1904.00945}}].

\bibitem{Badger:2019djh}
S.~Badger, D.~Chicherin, T.~Gehrmann, G.~Heinrich, J.~M. Henn, T.~Peraro,
  P.~Wasser, Y.~Zhang, and S.~Zoia, {\it {Analytic form of the full two-loop
  five-gluon all-plus helicity amplitude}},  {\em Phys. Rev. Lett.} {\bf 123}
  (2019), no.~7 071601, [\href{http://arxiv.org/abs/1905.03733}{{\tt
  arXiv:1905.03733}}].

\bibitem{Hartanto:2019uvl}
H.~B. Hartanto, S.~Badger, C.~Br\o{}nnum-Hansen, and T.~Peraro, {\it {A
  numerical evaluation of planar two-loop helicity amplitudes for a W-boson
  plus four partons}},  {\em JHEP} {\bf 09} (2019) 119,
  [\href{http://arxiv.org/abs/1906.11862}{{\tt arXiv:1906.11862}}].

\bibitem{Chawdhry:2020for}
H.~A. Chawdhry, M.~Czakon, A.~Mitov, and R.~Poncelet, {\it {Two-loop
  leading-color helicity amplitudes for three-photon production at the LHC}},
  {\em JHEP} {\bf 06} (2021) 150, [\href{http://arxiv.org/abs/2012.13553}{{\tt
  arXiv:2012.13553}}].

\bibitem{Kallweit:2020gcp}
S.~Kallweit, V.~Sotnikov, and M.~Wiesemann, {\it {Triphoton production at
  hadron colliders in NNLO QCD}},  {\em Phys. Lett. B} {\bf 812} (2021) 136013,
  [\href{http://arxiv.org/abs/2010.04681}{{\tt arXiv:2010.04681}}].

\bibitem{Abreu:2021oya}
S.~Abreu, F.~Febres~Cordero, H.~Ita, B.~Page, and V.~Sotnikov, {\it
  {Leading-color two-loop QCD corrections for three-jet production at hadron
  colliders}},  {\em JHEP} {\bf 07} (2021) 095,
  [\href{http://arxiv.org/abs/2102.13609}{{\tt arXiv:2102.13609}}].

\bibitem{Badger:2021nhg}
S.~Badger, H.~B. Hartanto, and S.~Zoia, {\it {Two-Loop QCD Corrections to
  Wbb\textasciimacron{} Production at Hadron Colliders}},  {\em Phys. Rev.
  Lett.} {\bf 127} (2021), no.~1 012001,
  [\href{http://arxiv.org/abs/2102.02516}{{\tt arXiv:2102.02516}}].

\bibitem{Badger:2021imn}
S.~Badger, C.~Br\o{}nnum-Hansen, D.~Chicherin, T.~Gehrmann, H.~B. Hartanto,
  J.~Henn, M.~Marcoli, R.~Moodie, T.~Peraro, and S.~Zoia, {\it {Virtual QCD
  corrections to gluon-initiated diphoton plus jet production at hadron
  colliders}},  {\em JHEP} {\bf 11} (2021) 083,
  [\href{http://arxiv.org/abs/2106.08664}{{\tt arXiv:2106.08664}}].

\bibitem{Badger:2021ega}
S.~Badger, H.~B. Hartanto, J.~Kry\'s, and S.~Zoia, {\it {Two-loop
  leading-colour QCD helicity amplitudes for Higgs boson production in
  association with a bottom-quark pair at the LHC}},  {\em JHEP} {\bf 11}
  (2021) 012, [\href{http://arxiv.org/abs/2107.14733}{{\tt arXiv:2107.14733}}].

\bibitem{Chawdhry:2021mkw}
H.~A. Chawdhry, M.~Czakon, A.~Mitov, and R.~Poncelet, {\it {Two-loop
  leading-colour QCD helicity amplitudes for two-photon plus jet production at
  the LHC}},  {\em JHEP} {\bf 07} (2021) 164,
  [\href{http://arxiv.org/abs/2103.04319}{{\tt arXiv:2103.04319}}].

\bibitem{Chawdhry:2021hkp}
H.~A. Chawdhry, M.~Czakon, A.~Mitov, and R.~Poncelet, {\it {NNLO QCD
  corrections to diphoton production with an additional jet at the LHC}},  {\em
  JHEP} {\bf 09} (2021) 093, [\href{http://arxiv.org/abs/2105.06940}{{\tt
  arXiv:2105.06940}}].

\bibitem{Czakon:2021mjy}
M.~Czakon, A.~Mitov, and R.~Poncelet, {\it {Next-to-Next-to-Leading Order Study
  of Three-Jet Production at the LHC}},  {\em Phys. Rev. Lett.} {\bf 127}
  (2021), no.~15 152001, [\href{http://arxiv.org/abs/2106.05331}{{\tt
  arXiv:2106.05331}}]. [Erratum: Phys.Rev.Lett. 129, 119901 (2022)].

\bibitem{Abreu:2021asb}
S.~Abreu, F.~Febres~Cordero, H.~Ita, M.~Klinkert, B.~Page, and V.~Sotnikov,
  {\it {Leading-color two-loop amplitudes for four partons and a W boson in
  QCD}},  {\em JHEP} {\bf 04} (2022) 042,
  [\href{http://arxiv.org/abs/2110.07541}{{\tt arXiv:2110.07541}}].

\bibitem{Badger:2022ncb}
S.~Badger, H.~B. Hartanto, J.~Kry\'s, and S.~Zoia, {\it {Two-loop leading
  colour helicity amplitudes for W$^{±}$\ensuremath{\gamma} + j production at
  the LHC}},  {\em JHEP} {\bf 05} (2022) 035,
  [\href{http://arxiv.org/abs/2201.04075}{{\tt arXiv:2201.04075}}].

\bibitem{Badger:2023mgf}
S.~Badger, M.~Czakon, H.~B. Hartanto, R.~Moodie, T.~Peraro, R.~Poncelet, and
  S.~Zoia, {\it {Isolated photon production in association with a jet pair
  through next-to-next-to-leading order in QCD}},  {\em JHEP} {\bf 10} (2023)
  071, [\href{http://arxiv.org/abs/2304.06682}{{\tt arXiv:2304.06682}}].

\bibitem{Abreu:2023bdp}
S.~Abreu, G.~De~Laurentis, H.~Ita, M.~Klinkert, B.~Page, and V.~Sotnikov, {\it
  {Two-loop QCD corrections for three-photon production at hadron colliders}},
  {\em SciPost Phys.} {\bf 15} (2023), no.~4 157,
  [\href{http://arxiv.org/abs/2305.17056}{{\tt arXiv:2305.17056}}].

\bibitem{Agarwal:2023suw}
B.~Agarwal, F.~Buccioni, F.~Devoto, G.~Gambuti, A.~von Manteuffel, and
  L.~Tancredi, {\it {Five-parton scattering in QCD at two loops}},  {\em Phys.
  Rev. D} {\bf 109} (2024), no.~9 094025,
  [\href{http://arxiv.org/abs/2311.09870}{{\tt arXiv:2311.09870}}].

\bibitem{DeLaurentis:2023nss}
G.~De~Laurentis, H.~Ita, M.~Klinkert, and V.~Sotnikov, {\it {Double-virtual
  NNLO QCD corrections for five-parton scattering. I. The gluon channel}},
  {\em Phys. Rev. D} {\bf 109} (2024), no.~9 094023,
  [\href{http://arxiv.org/abs/2311.10086}{{\tt arXiv:2311.10086}}].

\bibitem{DeLaurentis:2023izi}
G.~De~Laurentis, H.~Ita, and V.~Sotnikov, {\it {Double-virtual NNLO QCD
  corrections for five-parton scattering. II. The quark channels}},  {\em Phys.
  Rev. D} {\bf 109} (2024), no.~9 094024,
  [\href{http://arxiv.org/abs/2311.18752}{{\tt arXiv:2311.18752}}].

\bibitem{Badger:2024sqv}
S.~Badger, H.~B. Hartanto, Z.~Wu, Y.~Zhang, and S.~Zoia, {\it {Two-loop
  amplitudes for $ \mathcal{O}\left({\alpha}_s^2\right) $ corrections to
  W\ensuremath{\gamma}\ensuremath{\gamma} production at the LHC}},  {\em JHEP}
  {\bf 12} (2025) 221, [\href{http://arxiv.org/abs/2409.08146}{{\tt
  arXiv:2409.08146}}].

\bibitem{Badger:2024mir}
S.~Badger, H.~B. Hartanto, R.~Poncelet, Z.~Wu, Y.~Zhang, and S.~Zoia, {\it
  {Full-colour double-virtual amplitudes for associated production of a Higgs
  boson with a bottom-quark pair at the LHC}},  {\em JHEP} {\bf 03} (2025) 066,
  [\href{http://arxiv.org/abs/2412.06519}{{\tt arXiv:2412.06519}}].

\bibitem{Mazzitelli:2024ura}
J.~Mazzitelli, V.~Sotnikov, and M.~Wiesemann, {\it {Next-to-next-to-leading
  order event generation for Z-boson production in association with a
  bottom-quark pair}},  \href{http://arxiv.org/abs/2404.08598}{{\tt
  arXiv:2404.08598}}.

\bibitem{Badger:2024dxo}
S.~Badger, M.~Becchetti, C.~Brancaccio, H.~B. Hartanto, and S.~Zoia, {\it
  {Numerical evaluation of two-loop QCD helicity amplitudes for $ gg\to
  t\overline{t}g $ at leading colour}},  {\em JHEP} {\bf 03} (2025) 070,
  [\href{http://arxiv.org/abs/2412.13876}{{\tt arXiv:2412.13876}}].

\bibitem{Agarwal:2024jyq}
B.~Agarwal, G.~Heinrich, S.~P. Jones, M.~Kerner, S.~Y. Klein, J.~Lang,
  V.~Magerya, and A.~Olsson, {\it {Two-loop amplitudes for $ t\overline{t}H $
  production: the quark-initiated N$_{f}$-part}},  {\em JHEP} {\bf 05} (2024)
  013, [\href{http://arxiv.org/abs/2402.03301}{{\tt arXiv:2402.03301}}].
  [Erratum: JHEP 06, 142 (2024)].

\bibitem{DeLaurentis:2025dxw}
G.~De~Laurentis, H.~Ita, B.~Page, and V.~Sotnikov, {\it {Compact Two-Loop QCD
  Corrections for $Vjj$ Production in Proton Collisions}},
  \href{http://arxiv.org/abs/2503.10595}{{\tt arXiv:2503.10595}}.

\bibitem{Hahn:2000kx}
T.~Hahn, {\it {Generating Feynman diagrams and amplitudes with FeynArts 3}},
  {\em Comput. Phys. Commun.} {\bf 140} (2001) 418--431,
  [\href{http://arxiv.org/abs/hep-ph/0012260}{{\tt hep-ph/0012260}}].

\bibitem{Nogueira:1991ex}
P.~Nogueira, {\it {Automatic Feynman Graph Generation}},  {\em J. Comput.
  Phys.} {\bf 105} (1993) 279--289.

\bibitem{Pozzorini:2022ohr}
S.~Pozzorini, N.~Sch\"ar, and M.~F. Zoller, {\it {Two-loop tensor integral
  coefficients in OpenLoops}},  {\em JHEP} {\bf 05} (2022) 161,
  [\href{http://arxiv.org/abs/2201.11615}{{\tt arXiv:2201.11615}}].

\bibitem{Canko:2023lvh}
D.~Canko, G.~Bevilacqua, and C.~Papadopoulos, {\it {Two-Loop Amplitude
  Reduction with HELAC}},  {\em PoS} {\bf RADCOR2023} (2024) 081,
  [\href{http://arxiv.org/abs/2309.14886}{{\tt arXiv:2309.14886}}].

\bibitem{Peraro:2019cjj}
T.~Peraro and L.~Tancredi, {\it {Physical projectors for multi-leg helicity
  amplitudes}},  {\em JHEP} {\bf 07} (2019) 114,
  [\href{http://arxiv.org/abs/1906.03298}{{\tt arXiv:1906.03298}}].

\bibitem{Peraro:2020sfm}
T.~Peraro and L.~Tancredi, {\it {Tensor decomposition for bosonic and fermionic
  scattering amplitudes}},  {\em Phys. Rev. D} {\bf 103} (2021), no.~5 054042,
  [\href{http://arxiv.org/abs/2012.00820}{{\tt arXiv:2012.00820}}].

\bibitem{Anastasiou:2023koq}
C.~Anastasiou, J.~Karlen, and M.~Vicini, {\it {Tensor reduction of loop
  integrals}},  {\em JHEP} {\bf 12} (2023) 169,
  [\href{http://arxiv.org/abs/2308.14701}{{\tt arXiv:2308.14701}}].

\bibitem{Goode:2024mci}
J.~Goode, F.~Herzog, A.~Kennedy, S.~Teale, and J.~Vermaseren, {\it {Tensor
  reduction for Feynman integrals with Lorentz and spinor indices}},  {\em
  JHEP} {\bf 11} (2024) 123, [\href{http://arxiv.org/abs/2408.05137}{{\tt
  arXiv:2408.05137}}].

\bibitem{Goode:2024cfy}
J.~Goode, F.~Herzog, and S.~Teale, {\it {OPITeR: A program for tensor reduction
  of multi-loop Feynman integrals}},  {\em Comput. Phys. Commun.} {\bf 312}
  (2025) 109606, [\href{http://arxiv.org/abs/2411.02233}{{\tt
  arXiv:2411.02233}}].

\bibitem{Kosower:2011ty}
D.~A. Kosower and K.~J. Larsen, {\it {Maximal Unitarity at Two Loops}},  {\em
  Phys. Rev. D} {\bf 85} (2012) 045017,
  [\href{http://arxiv.org/abs/1108.1180}{{\tt arXiv:1108.1180}}].

\bibitem{Mastrolia:2011pr}
P.~Mastrolia and G.~Ossola, {\it {On the Integrand-Reduction Method for
  Two-Loop Scattering Amplitudes}},  {\em JHEP} {\bf 11} (2011) 014,
  [\href{http://arxiv.org/abs/1107.6041}{{\tt arXiv:1107.6041}}].

\bibitem{Zhang:2012ce}
Y.~Zhang, {\it {Integrand-Level Reduction of Loop Amplitudes by Computational
  Algebraic Geometry Methods}},  {\em JHEP} {\bf 09} (2012) 042,
  [\href{http://arxiv.org/abs/1205.5707}{{\tt arXiv:1205.5707}}].

\bibitem{Badger:2012dp}
S.~Badger, H.~Frellesvig, and Y.~Zhang, {\it {Hepta-Cuts of Two-Loop Scattering
  Amplitudes}},  {\em JHEP} {\bf 04} (2012) 055,
  [\href{http://arxiv.org/abs/1202.2019}{{\tt arXiv:1202.2019}}].

\bibitem{Mastrolia:2012an}
P.~Mastrolia, E.~Mirabella, G.~Ossola, and T.~Peraro, {\it {Scattering
  Amplitudes from Multivariate Polynomial Division}},  {\em Phys. Lett. B} {\bf
  718} (2012) 173--177, [\href{http://arxiv.org/abs/1205.7087}{{\tt
  arXiv:1205.7087}}].

\bibitem{Mastrolia:2012wf}
P.~Mastrolia, E.~Mirabella, G.~Ossola, and T.~Peraro, {\it {Integrand-Reduction
  for Two-Loop Scattering Amplitudes through Multivariate Polynomial
  Division}},  {\em Phys. Rev. D} {\bf 87} (2013), no.~8 085026,
  [\href{http://arxiv.org/abs/1209.4319}{{\tt arXiv:1209.4319}}].

\bibitem{Kleiss:2012yv}
R.~H.~P. Kleiss, I.~Malamos, C.~G. Papadopoulos, and R.~Verheyen, {\it
  {Counting to One: Reducibility of One- and Two-Loop Amplitudes at the
  Integrand Level}},  {\em JHEP} {\bf 12} (2012) 038,
  [\href{http://arxiv.org/abs/1206.4180}{{\tt arXiv:1206.4180}}].

\bibitem{Mastrolia:2013kca}
P.~Mastrolia, E.~Mirabella, G.~Ossola, and T.~Peraro, {\it {Multiloop Integrand
  Reduction for Dimensionally Regulated Amplitudes}},  {\em Phys. Lett. B} {\bf
  727} (2013) 532--535, [\href{http://arxiv.org/abs/1307.5832}{{\tt
  arXiv:1307.5832}}].

\bibitem{Badger:2013gxa}
S.~Badger, H.~Frellesvig, and Y.~Zhang, {\it {A Two-Loop Five-Gluon Helicity
  Amplitude in QCD}},  {\em JHEP} {\bf 12} (2013) 045,
  [\href{http://arxiv.org/abs/1310.1051}{{\tt arXiv:1310.1051}}].

\bibitem{Ita:2015tya}
H.~Ita, {\it {Two-loop Integrand Decomposition into Master Integrals and
  Surface Terms}},  {\em Phys. Rev. D} {\bf 94} (2016), no.~11 116015,
  [\href{http://arxiv.org/abs/1510.05626}{{\tt arXiv:1510.05626}}].

\bibitem{Mastrolia:2016dhn}
P.~Mastrolia, T.~Peraro, and A.~Primo, {\it {Adaptive Integrand Decomposition
  in parallel and orthogonal space}},  {\em JHEP} {\bf 08} (2016) 164,
  [\href{http://arxiv.org/abs/1605.03157}{{\tt arXiv:1605.03157}}].

\bibitem{Peraro:2019okx}
T.~Peraro, {\it {Analytic multi-loop results using finite fields and dataflow
  graphs with FiniteFlow}},  in {\em {14th International Symposium on Radiative
  Corrections}: {Application of Quantum Field Theory to Phenomenology}}, 12,
  2019.
\newblock \href{http://arxiv.org/abs/1912.03142}{{\tt arXiv:1912.03142}}.

\bibitem{Abreu_2021_C}
S.~Abreu, J.~Dormans, F.~Febres~Cordero, H.~Ita, M.~Kraus, B.~Page, E.~Pascual,
  M.~Ruf, and V.~Sotnikov, {\it Caravel: A c++ framework for the computation of
  multi-loop amplitudes with numerical unitarity},  {\em Computer Physics
  Communications} {\bf 267} (Oct., 2021) 108069.

\bibitem{Tkachov:1981wb}
F.~V. Tkachov, {\it {A Theorem on Analytical Calculability of Four Loop
  Renormalization Group Functions}},  {\em Phys. Lett.} {\bf 100B} (1981)
  65--68.

\bibitem{Chetyrkin:1981qh}
K.~G. Chetyrkin and F.~V. Tkachov, {\it {Integration by Parts: The Algorithm to
  Calculate beta Functions in 4 Loops}},  {\em Nucl. Phys. B} {\bf 192} (1981)
  159--204.

\bibitem{Laporta:2001dd}
S.~Laporta, {\it {High precision calculation of multiloop Feynman integrals by
  difference equations}},  {\em Int. J. Mod. Phys.} {\bf A15} (2000)
  5087--5159, [\href{http://arxiv.org/abs/hep-ph/0102033}{{\tt
  hep-ph/0102033}}].

\bibitem{vonManteuffel:2014ixa}
A.~von Manteuffel and R.~M. Schabinger, {\it {A novel approach to integration
  by parts reduction}},  {\em Phys. Lett. B} {\bf 744} (2015) 101--104,
  [\href{http://arxiv.org/abs/1406.4513}{{\tt arXiv:1406.4513}}].

\bibitem{Peraro:2016wsq}
T.~Peraro, {\it {Scattering amplitudes over finite fields and multivariate
  functional reconstruction}},  {\em JHEP} {\bf 12} (2016) 030,
  [\href{http://arxiv.org/abs/1608.01902}{{\tt arXiv:1608.01902}}].

\bibitem{Larsen:2015ped}
K.~J. Larsen and Y.~Zhang, {\it {Integration-by-parts reductions from unitarity
  cuts and algebraic geometry}},  {\em Phys. Rev. D} {\bf 93} (2016), no.~4
  041701, [\href{http://arxiv.org/abs/1511.01071}{{\tt arXiv:1511.01071}}].

\bibitem{Wu:2023upw}
Z.~Wu, J.~Boehm, R.~Ma, H.~Xu, and Y.~Zhang, {\it {NeatIBP 1.0, a package
  generating small-size integration-by-parts relations for Feynman integrals}},
   {\em Comput. Phys. Commun.} {\bf 295} (2024) 108999,
  [\href{http://arxiv.org/abs/2305.08783}{{\tt arXiv:2305.08783}}].

\bibitem{Liu:2018dmc}
X.~Liu and Y.-Q. Ma, {\it {Determining arbitrary Feynman integrals by vacuum
  integrals}},  {\em Phys. Rev. D} {\bf 99} (2019), no.~7 071501,
  [\href{http://arxiv.org/abs/1801.10523}{{\tt arXiv:1801.10523}}].

\bibitem{Guan:2024byi}
X.~Guan, X.~Liu, Y.-Q. Ma, and W.-H. Wu, {\it {Blade: A package for
  block-triangular form improved Feynman integrals decomposition}},  {\em
  Comput. Phys. Commun.} {\bf 310} (2025) 109538,
  [\href{http://arxiv.org/abs/2405.14621}{{\tt arXiv:2405.14621}}].

\bibitem{Binoth:1999sp}
T.~Binoth, J.~P. Guillet, and G.~Heinrich, {\it {Reduction formalism for
  dimensionally regulated one loop N point integrals}},  {\em Nucl. Phys. B}
  {\bf 572} (2000) 361--386, [\href{http://arxiv.org/abs/hep-ph/9911342}{{\tt
  hep-ph/9911342}}].

\bibitem{Heinrich:2008si}
G.~Heinrich, {\it {Sector Decomposition}},  {\em Int. J. Mod. Phys. A} {\bf 23}
  (2008) 1457--1486, [\href{http://arxiv.org/abs/0803.4177}{{\tt
  arXiv:0803.4177}}].

\bibitem{Heinrich:2023til}
G.~Heinrich, S.~P. Jones, M.~Kerner, V.~Magerya, A.~Olsson, and J.~Schlenk,
  {\it {Numerical scattering amplitudes with pySecDec}},  {\em Comput. Phys.
  Commun.} {\bf 295} (2024) 108956,
  [\href{http://arxiv.org/abs/2305.19768}{{\tt arXiv:2305.19768}}].

\bibitem{Barucchi:1973zm}
G.~Barucchi and G.~Ponzano, {\it {Differential equations for one-loop
  generalized Feynman integrals}},  {\em J. Math. Phys.} {\bf 14} (1973)
  396--401.

\bibitem{Kotikov:1990kg}
A.~V. Kotikov, {\it {Differential equations method: New technique for massive
  Feynman diagrams calculation}},  {\em Phys. Lett. B} {\bf 254} (1991)
  158--164.

\bibitem{Kotikov:1991hm}
A.~V. Kotikov, {\it {Differential equations method: The Calculation of vertex
  type Feynman diagrams}},  {\em Phys. Lett. B} {\bf 259} (1991) 314--322.

\bibitem{Gehrmann:1999as}
T.~Gehrmann and E.~Remiddi, {\it {Differential equations for two loop four
  point functions}},  {\em Nucl. Phys. B} {\bf 580} (2000) 485--518,
  [\href{http://arxiv.org/abs/hep-ph/9912329}{{\tt hep-ph/9912329}}].

\bibitem{Henn:2013pwa}
J.~M. Henn, {\it {Multiloop integrals in dimensional regularization made
  simple}},  {\em Phys. Rev. Lett.} {\bf 110} (2013) 251601,
  [\href{http://arxiv.org/abs/1304.1806}{{\tt arXiv:1304.1806}}].

\bibitem{Moriello:2019yhu}
F.~Moriello, {\it {Generalised power series expansions for the elliptic planar
  families of Higgs + jet production at two loops}},  {\em JHEP} {\bf 01}
  (2020) 150, [\href{http://arxiv.org/abs/1907.13234}{{\tt arXiv:1907.13234}}].

\bibitem{Liu:2017jxz}
X.~Liu, Y.-Q. Ma, and C.-Y. Wang, {\it {A Systematic and Efficient Method to
  Compute Multi-loop Master Integrals}},  {\em Phys. Lett. B} {\bf 779} (2018)
  353--357, [\href{http://arxiv.org/abs/1711.09572}{{\tt arXiv:1711.09572}}].

\bibitem{Liu:2022chg}
X.~Liu and Y.-Q. Ma, {\it {AMFlow: A Mathematica package for Feynman integrals
  computation via auxiliary mass flow}},  {\em Comput. Phys. Commun.} {\bf 283}
  (2023) 108565, [\href{http://arxiv.org/abs/2201.11669}{{\tt
  arXiv:2201.11669}}].

\bibitem{Huang:2024qan}
R.-J. Huang, D.-S. Jian, Y.-Q. Ma, D.-M. Mu, and W.-H. Wu, {\it {Efficient
  computation of one-loop Feynman integrals and fixed-branch integrals to high
  orders in \ensuremath{\varepsilon}}},  {\em Phys. Rev. D} {\bf 111} (2025),
  no.~9 094028, [\href{http://arxiv.org/abs/2412.21054}{{\tt
  arXiv:2412.21054}}].

\bibitem{Huang:2024nij}
L.-H. Huang, R.-J. Huang, and Y.-Q. Ma, {\it {Tame multi-leg Feynman integrals
  beyond one loop}},  \href{http://arxiv.org/abs/2412.21053}{{\tt
  arXiv:2412.21053}}.

\bibitem{Ossola:2006us}
G.~Ossola, C.~G. Papadopoulos, and R.~Pittau, {\it {Reducing full one-loop
  amplitudes to scalar integrals at the integrand level}},  {\em Nucl. Phys. B}
  {\bf 763} (2007) 147--169, [\href{http://arxiv.org/abs/hep-ph/0609007}{{\tt
  hep-ph/0609007}}].

\bibitem{Ossola:2007ax}
G.~Ossola, C.~G. Papadopoulos, and R.~Pittau, {\it {CutTools: A Program
  implementing the OPP reduction method to compute one-loop amplitudes}},  {\em
  JHEP} {\bf 03} (2008) 042, [\href{http://arxiv.org/abs/0711.3596}{{\tt
  arXiv:0711.3596}}].

\bibitem{Ellis:2007br}
R.~K. Ellis, W.~T. Giele, and Z.~Kunszt, {\it {A Numerical Unitarity Formalism
  for Evaluating One-Loop Amplitudes}},  {\em JHEP} {\bf 03} (2008) 003,
  [\href{http://arxiv.org/abs/0708.2398}{{\tt arXiv:0708.2398}}].

\bibitem{Ellis:2008ir}
R.~K. Ellis, W.~T. Giele, Z.~Kunszt, and K.~Melnikov, {\it {Masses, fermions
  and generalized $D$-dimensional unitarity}},  {\em Nucl. Phys. B} {\bf 822}
  (2009) 270--282, [\href{http://arxiv.org/abs/0806.3467}{{\tt
  arXiv:0806.3467}}].

\bibitem{Berger:2008sj}
C.~F. Berger, Z.~Bern, L.~J. Dixon, F.~Febres~Cordero, D.~Forde, H.~Ita, D.~A.
  Kosower, and D.~Maitre, {\it {An Automated Implementation of On-Shell Methods
  for One-Loop Amplitudes}},  {\em Phys. Rev. D} {\bf 78} (2008) 036003,
  [\href{http://arxiv.org/abs/0803.4180}{{\tt arXiv:0803.4180}}].

\bibitem{vanHameren:2009dr}
A.~van Hameren, C.~G. Papadopoulos, and R.~Pittau, {\it {Automated one-loop
  calculations: A Proof of concept}},  {\em JHEP} {\bf 09} (2009) 106,
  [\href{http://arxiv.org/abs/0903.4665}{{\tt arXiv:0903.4665}}].

\bibitem{vanHameren:2009vq}
A.~van Hameren, {\it {Multi-gluon one-loop amplitudes using tensor integrals}},
   {\em JHEP} {\bf 07} (2009) 088, [\href{http://arxiv.org/abs/0905.1005}{{\tt
  arXiv:0905.1005}}].

\bibitem{Mastrolia:2010nb}
P.~Mastrolia, G.~Ossola, T.~Reiter, and F.~Tramontano, {\it {Scattering
  AMplitudes from Unitarity-based Reduction Algorithm at the Integrand-level}},
   {\em JHEP} {\bf 08} (2010) 080, [\href{http://arxiv.org/abs/1006.0710}{{\tt
  arXiv:1006.0710}}].

\bibitem{Badger:2010nx}
S.~Badger, B.~Biedermann, and P.~Uwer, {\it {NGluon: A Package to Calculate
  One-loop Multi-gluon Amplitudes}},  {\em Comput. Phys. Commun.} {\bf 182}
  (2011) 1674--1692, [\href{http://arxiv.org/abs/1011.2900}{{\tt
  arXiv:1011.2900}}].

\bibitem{Bevilacqua:2011xh}
G.~Bevilacqua, M.~Czakon, M.~V. Garzelli, A.~van Hameren, A.~Kardos, C.~G.
  Papadopoulos, R.~Pittau, and M.~Worek, {\it {HELAC-NLO}},  {\em Comput. Phys.
  Commun.} {\bf 184} (2013) 986--997,
  [\href{http://arxiv.org/abs/1110.1499}{{\tt arXiv:1110.1499}}].

\bibitem{Peraro:2014cba}
T.~Peraro, {\it {Ninja: Automated Integrand Reduction via Laurent Expansion for
  One-Loop Amplitudes}},  {\em Comput. Phys. Commun.} {\bf 185} (2014)
  2771--2797, [\href{http://arxiv.org/abs/1403.1229}{{\tt arXiv:1403.1229}}].

\bibitem{Alwall:2014hca}
J.~Alwall, R.~Frederix, S.~Frixione, V.~Hirschi, F.~Maltoni, O.~Mattelaer,
  H.~S. Shao, T.~Stelzer, P.~Torrielli, and M.~Zaro, {\it {The automated
  computation of tree-level and next-to-leading order differential cross
  sections, and their matching to parton shower simulations}},  {\em JHEP} {\bf
  07} (2014) 079, [\href{http://arxiv.org/abs/1405.0301}{{\tt
  arXiv:1405.0301}}].

\bibitem{GoSam:2014iqq}
{\bf GoSam} Collaboration, G.~Cullen et~al., {\it
  {G$\scriptsize{O}$S$\scriptsize{AM}$-2.0: a tool for automated one-loop
  calculations within the Standard Model and beyond}},  {\em Eur. Phys. J. C}
  {\bf 74} (2014), no.~8 3001, [\href{http://arxiv.org/abs/1404.7096}{{\tt
  arXiv:1404.7096}}].

\bibitem{Buccioni:2019sur}
F.~Buccioni, J.-N. Lang, J.~M. Lindert, P.~Maierh\"ofer, S.~Pozzorini,
  H.~Zhang, and M.~F. Zoller, {\it {OpenLoops 2}},  {\em Eur. Phys. J. C} {\bf
  79} (2019), no.~10 866, [\href{http://arxiv.org/abs/1907.13071}{{\tt
  arXiv:1907.13071}}].

\bibitem{Bevilacqua:2024fec}
G.~Bevilacqua, D.~Canko, C.~G. Papadopoulos, and A.~Spourdalakis, {\it
  {Two-loop amplitude computation with HELAC}},  {\em PoS} {\bf LL2024} (2024)
  051.

\bibitem{Sotnikov:2019onv}
V.~Sotnikov, {\em {Scattering amplitudes with the multi-loop numerical
  unitarity method}}.
\newblock PhD thesis, Freiburg U., Freiburg U., 9, 2019.

\bibitem{Ossola:2008xq}
G.~Ossola, C.~G. Papadopoulos, and R.~Pittau, {\it {On the Rational Terms of
  the one-loop amplitudes}},  {\em JHEP} {\bf 05} (2008) 004,
  [\href{http://arxiv.org/abs/0802.1876}{{\tt arXiv:0802.1876}}].

\bibitem{Badger:2008cm}
S.~D. Badger, {\it {Direct Extraction Of One Loop Rational Terms}},  {\em JHEP}
  {\bf 01} (2009) 049, [\href{http://arxiv.org/abs/0806.4600}{{\tt
  arXiv:0806.4600}}].

\bibitem{Pozzorini:2020hkx}
S.~Pozzorini, H.~Zhang, and M.~F. Zoller, {\it {Rational Terms of UV Origin at
  Two Loops}},  {\em JHEP} {\bf 05} (2020) 077,
  [\href{http://arxiv.org/abs/2001.11388}{{\tt arXiv:2001.11388}}].

\bibitem{Lang:2020nnl}
J.-N. Lang, S.~Pozzorini, H.~Zhang, and M.~F. Zoller, {\it {Two-Loop Rational
  Terms in Yang-Mills Theories}},  {\em JHEP} {\bf 10} (2020) 016,
  [\href{http://arxiv.org/abs/2007.03713}{{\tt arXiv:2007.03713}}].

\bibitem{Lang:2021hnw}
J.-N. Lang, S.~Pozzorini, H.~Zhang, and M.~F. Zoller, {\it {Two-loop rational
  terms for spontaneously broken theories}},  {\em JHEP} {\bf 01} (2022) 105,
  [\href{http://arxiv.org/abs/2107.10288}{{\tt arXiv:2107.10288}}].

\bibitem{eigenweb}
G.~Guennebaud, B.~Jacob, et~al., ``Eigen v3.'' http://eigen.tuxfamily.org,
  2010.

\bibitem{LaPack}
E.~Anderson, Z.~Bai, C.~Bischof, S.~Blackford, J.~Demmel, J.~Dongarra,
  J.~Du~Croz, A.~Greenbaum, S.~Hammarling, A.~McKenney, and D.~Sorensen, {\em
  {LAPACK} Users' Guide}.
\newblock Society for Industrial and Applied Mathematics, Philadelphia, PA,
  third~ed., 1999.

\bibitem{Shtabovenko:2023idz}
V.~Shtabovenko, R.~Mertig, and F.~Orellana, {\it {FeynCalc 10: Do multiloop
  integrals dream of computer codes?}},  {\em Comput. Phys. Commun.} {\bf 306}
  (2025) 109357, [\href{http://arxiv.org/abs/2312.14089}{{\tt
  arXiv:2312.14089}}].

\bibitem{Ruijl:2017dtg}
B.~Ruijl, T.~Ueda, and J.~Vermaseren, {\it {FORM version 4.2}},
  \href{http://arxiv.org/abs/1707.06453}{{\tt arXiv:1707.06453}}.

\bibitem{Abreu:2024fei}
S.~Abreu, P.~F. Monni, B.~Page, and J.~Usovitsch, {\it {Planar Six-Point
  Feynman Integrals for Four-Dimensional Gauge Theories}},
  \href{http://arxiv.org/abs/2412.19884}{{\tt arXiv:2412.19884}}.

\bibitem{new1}
G.~Bevilacqua, D.~Canko, C.~G. Papadopoulos, and A.~Spourdalakis, {\it
  {Numerical evaluation of dimensionally regulated amplitudes}},  {\em {\rm
  work in progress}}.

\bibitem{Papadopoulos:2015jft}
C.~G. Papadopoulos, D.~Tommasini, and C.~Wever, {\it {The Pentabox Master
  Integrals with the Simplified Differential Equations approach}},  {\em JHEP}
  {\bf 04} (2016) 078, [\href{http://arxiv.org/abs/1511.09404}{{\tt
  arXiv:1511.09404}}].

\bibitem{Gehrmann:2018yef}
T.~Gehrmann, J.~M. Henn, and N.~A. Lo~Presti, {\it {Pentagon functions for
  massless planar scattering amplitudes}},  {\em JHEP} {\bf 10} (2018) 103,
  [\href{http://arxiv.org/abs/1807.09812}{{\tt arXiv:1807.09812}}].

\bibitem{Chicherin:2018mue}
D.~Chicherin, T.~Gehrmann, J.~M. Henn, N.~A. Lo~Presti, V.~Mitev, and
  P.~Wasser, {\it {Analytic result for the nonplanar hexa-box integrals}},
  {\em JHEP} {\bf 03} (2019) 042, [\href{http://arxiv.org/abs/1809.06240}{{\tt
  arXiv:1809.06240}}].

\bibitem{Chicherin:2018old}
D.~Chicherin, T.~Gehrmann, J.~M. Henn, P.~Wasser, Y.~Zhang, and S.~Zoia, {\it
  {All Master Integrals for Three-Jet Production at Next-to-Next-to-Leading
  Order}},  {\em Phys. Rev. Lett.} {\bf 123} (2019), no.~4 041603,
  [\href{http://arxiv.org/abs/1812.11160}{{\tt arXiv:1812.11160}}].

\bibitem{Abreu:2020jxa}
S.~Abreu, H.~Ita, F.~Moriello, B.~Page, W.~Tschernow, and M.~Zeng, {\it
  {Two-Loop Integrals for Planar Five-Point One-Mass Processes}},  {\em JHEP}
  {\bf 11} (2020) 117, [\href{http://arxiv.org/abs/2005.04195}{{\tt
  arXiv:2005.04195}}].

\bibitem{Canko:2020ylt}
D.~D. Canko, C.~G. Papadopoulos, and N.~Syrrakos, {\it {Analytic representation
  of all planar two-loop five-point Master Integrals with one off-shell leg}},
  {\em JHEP} {\bf 01} (2021) 199, [\href{http://arxiv.org/abs/2009.13917}{{\tt
  arXiv:2009.13917}}].

\bibitem{Chicherin:2020oor}
D.~Chicherin and V.~Sotnikov, {\it {Pentagon Functions for Scattering of Five
  Massless Particles}},  {\em JHEP} {\bf 20} (2020) 167,
  [\href{http://arxiv.org/abs/2009.07803}{{\tt arXiv:2009.07803}}].

\bibitem{Abreu:2021smk}
S.~Abreu, H.~Ita, B.~Page, and W.~Tschernow, {\it {Two-loop hexa-box integrals
  for non-planar five-point one-mass processes}},  {\em JHEP} {\bf 03} (2022)
  182, [\href{http://arxiv.org/abs/2107.14180}{{\tt arXiv:2107.14180}}].

\bibitem{Chicherin:2021dyp}
D.~Chicherin, V.~Sotnikov, and S.~Zoia, {\it {Pentagon functions for one-mass
  planar scattering amplitudes}},  {\em JHEP} {\bf 01} (2022) 096,
  [\href{http://arxiv.org/abs/2110.10111}{{\tt arXiv:2110.10111}}].

\bibitem{Kardos:2022tpo}
A.~Kardos, C.~G. Papadopoulos, A.~V. Smirnov, N.~Syrrakos, and C.~Wever, {\it
  {Two-loop non-planar hexa-box integrals with one massive leg}},  {\em JHEP}
  {\bf 05} (2022) 033, [\href{http://arxiv.org/abs/2201.07509}{{\tt
  arXiv:2201.07509}}].

\bibitem{Badger:2022hno}
S.~Badger, M.~Becchetti, E.~Chaubey, and R.~Marzucca, {\it {Two-loop master
  integrals for a planar topology contributing to pp \textrightarrow{}$
  t\overline{t}j $}},  {\em JHEP} {\bf 01} (2023) 156,
  [\href{http://arxiv.org/abs/2210.17477}{{\tt arXiv:2210.17477}}].

\bibitem{FebresCordero:2023pww}
F.~Febres~Cordero, G.~Figueiredo, M.~Kraus, B.~Page, and L.~Reina, {\it
  {Two-loop master integrals for leading-color $ pp\to t\overline{t}H $
  amplitudes with a light-quark loop}},  {\em JHEP} {\bf 07} (2024) 084,
  [\href{http://arxiv.org/abs/2312.08131}{{\tt arXiv:2312.08131}}].

\bibitem{Abreu:2023rco}
S.~Abreu, D.~Chicherin, H.~Ita, B.~Page, V.~Sotnikov, W.~Tschernow, and
  S.~Zoia, {\it {All Two-Loop Feynman Integrals for Five-Point One-Mass
  Scattering}},  {\em Phys. Rev. Lett.} {\bf 132} (2024), no.~14 141601,
  [\href{http://arxiv.org/abs/2306.15431}{{\tt arXiv:2306.15431}}].

\bibitem{Abreu:2024yit}
S.~Abreu, D.~Chicherin, V.~Sotnikov, and S.~Zoia, {\it {Two-loop five-point
  two-mass planar integrals and double Lagrangian insertions in a Wilson
  loop}},  {\em JHEP} {\bf 10} (2024) 167,
  [\href{http://arxiv.org/abs/2408.05201}{{\tt arXiv:2408.05201}}].

\bibitem{Badger:2024fgb}
S.~Badger, M.~Becchetti, N.~Giraudo, and S.~Zoia, {\it {Two-loop integrals for
  $ t\overline{t} $+jet production at hadron colliders in the leading colour
  approximation}},  {\em JHEP} {\bf 07} (2024) 073,
  [\href{http://arxiv.org/abs/2404.12325}{{\tt arXiv:2404.12325}}].

\bibitem{Henn:2025xrc}
J.~Henn, A.~Matija\v{s}i\'c, J.~Miczajka, T.~Peraro, Y.~Xu, and Y.~Zhang, {\it
  {Complete function space for planar two-loop six-particle scattering
  amplitudes}},  \href{http://arxiv.org/abs/2501.01847}{{\tt
  arXiv:2501.01847}}.

\bibitem{Becchetti:2025qlu}
M.~Becchetti, D.~Canko, V.~Chestnov, T.~Peraro, M.~Pozzoli, and S.~Zoia, {\it
  {Two-loop Feynman integrals for leading colour $t\bar{t}W$ production at
  hadron colliders}},  \href{http://arxiv.org/abs/2504.13011}{{\tt
  arXiv:2504.13011}}.

\bibitem{Bargiela:2025nqc}
P.~Bargiela and T.-Z. Yang, {\it {On the finite basis of two-loop `t
  Hooft-Veltman Feynman integrals}},
  \href{http://arxiv.org/abs/2503.16299}{{\tt arXiv:2503.16299}}.

\bibitem{new2}
G.~Bevilacqua, D.~Canko, C.~G. Papadopoulos, and A.~Spourdalakis, {\it
  {Automated two-loop calculations: a proof of concept}},  {\em {\rm work in
  progress}}.

\end{thebibliography}\endgroup

\end{document}